\documentclass[12pt]{article}
\usepackage{amsmath,amssymb,amsfonts}
\usepackage[dvips]{graphicx}
\usepackage{epsfig}
\usepackage{cite}

\makeatletter \@addtoreset{equation}{section} \makeatother
\makeatletter \@addtoreset{figure}{section} \makeatother

\def\CA{{\cal A}}\def\CB{{\cal B}}\def\CD{{\cal D}}
\def\CE{{\cal E}}
\def\CH{{\cal H}}
\def\CK{{\cal K}}\def\CL{{\cal L}}
\def\CN{{\cal N}}
\def\CQ{{\cal Q}}

\def\CW{{\cal W}}
\def\CZ{{\cal Z}}

\def\a{\alpha}
\def\d{\delta}\def\e{\epsilon}
\def\z{\zeta}\def\th{\theta}
\def\l{\lambda}

\def\r{\rho}\def\s{\sigma}
\def\t{\tau}
\def\w{\omega}\def\G{\Gamma}
\def\D{\Delta}\def\L{\Lambda}
\def\S{\Sigma}

\def\bj{{\bar j}}
\def\bl{{\bar l}}\def\bm{{\bar m}}
\def\bn{{\bar n}}

\def\tr{{\rm tr}}

\def\da{{\dot{\a}}}

\newcommand{\nn}{\nonumber}

\begin{document}
\begin{titlepage}
\vfill
\begin{flushright}
{\tt\normalsize KIAS-P09052}\\
\end{flushright}
\vfill
\begin{center}
{\Large\bf A Study of Wall-Crossing: \\ \vskip 3mm Flavored Kinks in $D=2$ QED
}

\vfill

Sungjay Lee\footnote{\tt sjlee@kias.re.kr}
and Piljin Yi\footnote{\tt piljin@kias.re.kr}

\vskip 5mm
{\it  School of Physics, Korea Institute for Advanced Study, Seoul 130-722, Korea}

\end{center}
\vfill

\begin{abstract}
\noindent
We study spectrum of $D=2$ ${\cal N}=(2,2)$ QED with $N+1$ massive charged
chiral multiplets, with care given to precise supermultiplet countings.
In the infrared the theory flows to $\mathbb{CP}^N$ model with twisted masses,
where we construct generic flavored kink solitons for the large mass regime,
and study their quantum degeneracies. These kinks are qualitatively
different and far more numerous than those of small mass regime, with
features reminiscent of multi-pronged $(p,q)$ string web, complete
with the wall-crossing behavior. It has been also conjectured that spectrum
of this theory is equivalent to the hypermultiplet spectrum of a certain
$D=4$ Seiberg-Witten theory. We find that the correspondence
actually extends beyond hypermultiplets in $D=4$, and that many of
the relevant indices match. However, a $D=2$ BPS state is typically mapped to
several different kind of dyons whose individual supermultiplets are
rather complicated; the match of index comes about only after summing
over indices of these different dyons. We note general wall-crossing
behavior of flavored BPS kink states, and compare it to those of
$D=4$ dyons.

\end{abstract}

\vfill
\end{titlepage}

\tableofcontents\newpage
\renewcommand{\thefootnote}{\#\arabic{footnote}}
\setcounter{footnote}{0}

\section{Introduction}

The  wall-crossing in four-dimensional supersymmetric theory
\cite{Seiberg:1994rs, Ferrari:1996sv, Bilal:1996sk} has been
a subject of interests to many string theorists and mathematicians.
This phenomenon of discontinuity in the BPS spectrum across walls
of marginal stability, as one changes
either parameters or vacuum expectation value of a theory, has been
a source of enormous difficulty in understanding the detailed structure
of theories like $\CN=2$ Seiberg-Witten theories and Calabi-Yau
compactified type II string theories.

For BPS states in four-dimensional
theories, this phenomenon has been understood in various physical viewpoints,\footnote{See
Ref.~\cite{Weinberg:2006rq} for a review in the field theory side.}
such as from geometric realization of BPS states in string theory
\cite{Mikhailov:1998bx,Bergman:1997yw}, from solitonic dynamics
\cite{Lee:1998nv,Bak:1999da} and quantum bound states thereof
\cite{Bak:1999ip,Gauntlett:1999vc},
from a classical soliton picture of the low energy effective theory
\cite{Ritz:2000xa,Argyres:2001pv}, and also later from supergravity attractor
flow \cite{Denef:2000nb}. From the spacetime viewpoint, the wall-crossing
occurs simply because the wavefunction of the BPS state in question becomes
so large (as one approaches a wall of marginal stability) that the state
in question cannot be regarded as a one-particle state anymore
\cite{Bak:1999ip,Gauntlett:1999vc,Denef:2000nb}.
Despite this simple and compelling physical picture, a systematic
and practical approach to the wall-crossing phenomenon which
can cover all part of the moduli space had not been available.

Recently, there appeared a new remarkable development in this regard. It states that
such discontinuities of spectrum across  walls of marginal stability
is actually necessary for the continuity of
the vacuum moduli space metric. According to
Gaiotto, Moore and Neitzke (GMN) \cite{Gaiotto:2008cd,Gaiotto:2009hg},
the continuity of the vacuum moduli space
metric of $S^1$-compactified Seiberg-Witten theory implies the so-called
Kontsevich-Soibelman relations \cite{Kontsevich:2008ab}
among BPS dyons across any given wall of marginal
stability, which in turn tells us how the BPS spectra would change across
such walls. Cecotti and Vafa \cite{Cecotti:2009uf} has recently suggested
another interesting explanation of Kontsevich-Soibelman's formulae with
spin refinement \cite{Dimofte:2009bv}, using the partition function of
A-model topological string.

While the derivation by GMN was intended
for $\CN=2$ Seiberg-Witten theory, the idea
itself must be applicable to all wall-crossing phenomena. This new machinary
is also important in that for the first time we have a systematic and
local prescription for computing BPS spectrum. Although there were
powerful methods which allowed explicit construction/counting of BPS
states in certain regions of the moduli space
\cite{Bak:1999ip,Gauntlett:1999vc,Stern:2000ie,Denef:2002ru},
this new wall-crossing formula is far more comprehensive in its
potential applications.

This observation that discontinuity of BPS spectra is related to
continuity of some physical quantity has, on the other hand,
a previously known analog in the
context of two-dimensional $\CN=(2,2)$ theories. Cecotti and Vafa
\cite{Cecotti:1992qh,Cecotti:1992rm} noted some time
ago that if one assumes continuity of a twisted partition function
\begin{equation}
{\cal F}(\beta;m^i)=\tr (-1)^R R e^{-\beta H}
\end{equation}
throughout parameter space of the theory, this necessarily implies
(dis-)appearance of BPS topological kinks across walls of marginal
stability. Here $R$ is the fermion number, and $m^i$'s are the
parameters of the theory. The above twisted partition is in turn
related to the natural metric in the parameter space, and obeys
the so-called $tt^*$ equation \cite{Cecotti:1991me}. In fact,
GMN also noted that some of
mathematical structures of $tt^*$ equation is very closely
mirrored by those that appear in their formulation of the
four-dimensional wall-crossing.

Independent of this, another interesting similarity between $D=4$ $\CN=2$
and $D=2$ $\CN=(2,2)$ theories was noted in the literature:  It has been
conjectured  \cite{Dorey:1998yh,Dorey:1999zk}
that two-dimensional $\CN=2$ QED with $N+1$ massive chiral multiplets
possesses a BPS spectrum which is related to that of $SU(N+1)$
Seiberg-Witten theory with $N+1$ massive flavors at the root of the baryonic
branch.
In view of the new development in the Seiberg-Witten theory concerning
the wall-crossing, and given its analog in $tt^*$ system, it is
of some interest to clarify the precise correspondence and potential
differences. In this article, we aim to study the two-dimensional
theory with  care given to precise BPS multiplet countings, and compare
their wall-crossing phenomena against that of the Seiberg-Witten theory.

$\CN=(2,2)$ QED with $N+1$ chiral multiplets with twisted masses has been studied
much previously. Initial studies by Hanany and Hori \cite{Hanany:1997vm}
and also by Dorey \cite{Dorey:1998yh, Dorey:1999zk} concentrated on
implications of effective superpotential of the gauge-multiplet and
its similarity to certain Seiberg-Witten spectral curve of $D=4$ theory.
Later works \cite{Hanany:2004ea,Shifman:2004dr,Tong:2005un,Shifman:2007ce,Tong:2008qd}
refined this relationship further by giving physical reasonings,
if somewhat sketchy, for the correspondence and also looked at
$D=2$ spectrum more closely by considering massive excitations
of simple kink solutions.

In this paper, we expand on these existing works and solve for
all possible flavored kinks. We give precise criteria
for existence of such flavored kink states, set up the low energy
dynamics of kinks, count their degeneracies, and provide wall-crossing
formula. This allows a more refined look at the proposed
``equivalence'' of the spectra. We also hope that it will provide
a playground for understanding wall-crossing phenomena in $D=2$
when conserved charged other than the topological ones are present.

In section 2 and 3, we review the theory and search for all possible
kink soliton solutions. Although kinks are simple and well-known objects,
global charge allows the variety of kink solutions to increase greatly.
Apart from simple ``dyonic'' kinks whose flavor charge is proportional
to the topological charge, there are much more flavored kinks whose
central charges and stability criteria mimics those of the $(p,q)$
open string webs \cite{Bergman:1997yw,Lee:1998nv}. In section 4, we quantize these
solitons, elevate them to quantum BPS states, and count their degeneracy.
These BPS states exhibit wall-crossing behavior, just as open string web
does, which we put in the context of general $D=2$ and $\CN=(2,2)$ theories
following Cecotti and Vafa's results. In section 5, we compare this
spectra to its conjectured counterpart in $D=4$ Seiberg-Witten theory.
Although, the two sides have some common features, essentially due to
the open string web analogy, absence of ``angular momentum'' in the $D=2$
theory leads to quantitatively different spectra.  However,
a set of distinct dyons with different quark contents are mapped to
a single type of favored kink; interestingly, if one sum over the relevant indices
of the former, the result matches precisely with the degeneracy of
the flavored kink. We rely on the four-dimensional wall-crossing
formula to reach this conclusion. We close with conclusion.

\section{$\mathbb{CP}^N$ with twisted masses}

Let us first summarize basic properties of $\CN = (2,2)$ supersymmetric
theories in two dimensions.\footnote{Please see Appendix A  for fruther notations and conventions. }
In particular we discuss the massive representation
of $\CN = (2,2)$ SUSY algebra and the CFIV index \cite{Cecotti:1992qh}
which effectively counts the short multiplets only.

\paragraph{supersymmetry algebra}

The $\CN=(2,2)$ superalgebra can read off from
the four-dimensional $\CN=1$ superalgebra via
trivial dimensional reduction
as
\begin{eqnarray}\label{SUSY22}
  \big\{ Q_+ ,\bar Q_+ \big\} = 2 Z, &&
  \big\{ Q_+ , \bar Q_- \big\} = -2 \big( P_0 - P_3 \big),
  \nn \\
  \big\{ Q_- , \bar Q_- \big\} = 2 \bar Z,
  &&
  \big\{ Q_- , \bar Q_+ \big\} = -2 \big( P_0 + P_3 \big)\ ,
\end{eqnarray}
where the central charge $Z$ is
\begin{eqnarray}
  Z = P_1 - i P_2 \ .
\end{eqnarray}
For later convenience, let us summarize the $U(1)_\text{R}\times U(1)_\text{A}$
charges of supersymmetric generators
\begin{eqnarray}
  \begin{array}{c|cccc}
  & Q_+ & Q_- & \bar Q_+ & \bar Q_- \\
  \hline
  U(1)_\text{R} & +1 & +1 & -1 & -1 \\
  U(1)_\text{A} & +1 & -1 & +1 & -1
  \end{array}\ .
\end{eqnarray}
Here $U(1)_\text{A}$ symmetry comes from the rotational symmetry $SO(2)$ in
four dimensions.


In massive theories, one of the two $U(1)$ symmetries are explicitly
broken, and suppose we choose the following basis that preserveq
$U(1)_\text{R}$ 
%
\begin{eqnarray}
  \CA = \frac{1}{\sqrt2} \big( Q_+ + Q_- \big)\ , \qquad
  \CB = \frac{1}{\sqrt2} \big( Q_+ - Q_- \big)\ .
\end{eqnarray}
Making the central charge $Z$ real via a suitable $U(1)_\text{A}$
rotation, the supersymmetry algebra can be recast as
\begin{eqnarray}
  \big\{ \CA , \CA^\dagger \big\} = - 2 \big( M - Z \big)\ ,
  \qquad
  \big\{ \CB , \CB^\dagger \big\} = - 2 \big( M + Z \big)\ ,
  \qquad
  \big\{ \CA , \CB^\dagger \big\} = 0\ ,
\end{eqnarray}
One can therefore conclude that, for massive BPS multiplets, the algebra
eventually is reduced to that of a single fermion
oscillator.

\paragraph{CFIV index}
With this, the index that count BPS multiplets is
\begin{eqnarray}
  \Omega = \text{tr} \Big[ (-1)^R R \Big]\ .
\end{eqnarray}
This is a proper index since for long multiplets
in Fock vacuum of R-charge $f$
  \begin{eqnarray}
    [{f }] \otimes \big( [{\bf 1}] \oplus  [{\bf 0}] \big)^2
    \ \  \Longrightarrow \ \
    [{f+2 }] \oplus 2 [{ f+1 }] \oplus  [{f}]\ , \nonumber
  \end{eqnarray}
  the index $\Omega$ identically vanishes
  \begin{eqnarray}
    \Omega = 0\ .
  \end{eqnarray}
On the other hand, for generic BPS multiplets
  \begin{eqnarray}
    [{f}] \otimes \big( [{\bf 1}] \oplus [{\bf 0}] \big) \ \
    \Longrightarrow \  \
    [{ f+1 }] \oplus [{ f}]\ , \nonumber
  \end{eqnarray}
one can have non-vanishing $\Omega$
  \begin{eqnarray}
    \Omega = (-1)^{f+1} \ .
  \end{eqnarray}
The simplicity of $D=2$ theory is such that we have only
two types of BPS multiplets, labeled by this sign, which
is because of the small supersymmetry compounded by absence of
spin.\footnote{
The mirror symmetry, or t-duality in two-dimensional
supersymmetric theory, exchanges those two R-symmetries
\begin{eqnarray}
  U(1)_\text{R} \leftrightarrow U(1)_\text{A}\ , \qquad
  Q_- \leftrightarrow \bar Q_+\ .
\end{eqnarray}
In the mirror-symmetric dual, the proper index now in turn
is defined with $U(1)_\text{A}$ charge,
\begin{eqnarray}
  \Omega = \text{tr} \Big[ (-1)^A A \Big]\ . \nonumber
\end{eqnarray}
}

\subsection{review on massive $\mathbb{CP}^N$-model}

We consider a two-dimensional supersymmetric QED
which flows down to a massive $\mathbb{CP}^N$-model
with twisted masses. It is well-known that
the massless $\mathbb{CP}^N$-model can be easily
understood as IR limit of a gauged linear sigma
model (GLSM) with a photon field $V$
and $N+1$ chiral matter fields $\phi^i$ of unit charge.
Introducing the Fayet-Iliopoulos (FI) parameter $r$ together
with theta-angle $\th$, the Lagrangian takes the following form
\begin{eqnarray}\label{GLSMlag}
  \CL = \int d^4\th \ \Big[\phi^\dagger_i e^{-2V} \phi^i - \frac{1}{4e^2} \bar \Sigma
  \Sigma \Big] - \text{Im}\Big[ \t  \int d^2 \hat \theta \ \S \Big] \ ,
  \qquad \t = - i r + \frac{\th}{2\pi}\ ,
\end{eqnarray}
where $i$ run from $0,1,..,N$. Again, the notations and conventions
used here are introduced in appendix A.
For a positive FI parameter $r>0$, the
supersymmetric vacuum can be described by
\begin{eqnarray}
  \sum_i |\phi^i|^2 = r \ , \qquad \sigma= 0\ ,
\end{eqnarray}
which defines a projective space $\mathbb{CP}^N$. On the generic
point of vacuum moduli space, the $U(1)$ vector multiplet and
chiral mode orthogonal to $\mathbb{CP}^N$ are
combined to a long multiplet of mass $\sqrt r e$ by the Higgs mechanism.
In the IR limit where $e^2$ diverges, these modes become very heavy
so that they decouple from the low-energy dynamics of the theory.
It leads to a $\CN=(2,2)$ $\mathbb{CP}^N$ model.

We will present a simple way to obtain the effective Lagrangian
for the above low-energy theory, $\CN=(2,2)$ $\mathbb{CP}^N$ model.
For simplicity, let us first turn off the theta-angle $\th=0$ for a while.
Note that we can then rewrite the Fayet-Iliopoulos (FI) term as
\begin{eqnarray}
  \CL_\text{FI} =   2 r \int d^4\th \  V  \ .
\end{eqnarray}
The decoupling phenomenon of massive modes in the Higgs phase
can be realized effectively as the vanishing Maxwell term
in the limit of $e^2 \to \infty$.
The low-energy theory at IR is now governed by the following Lagrangian
\begin{eqnarray}
  \CL \simeq \int d^4\th \ \Big[\phi^\dagger_i e^{-2V} \phi^i + 2r V  \Big]\ .
\end{eqnarray}
Here the vector multiplet becomes an auxiliary fields that one can solve out:
\begin{eqnarray}
  \d V \ : \ \  r =  \phi^\dagger_i e^{-2V} \phi^i \ \Rightarrow \
  V = -\frac12 \text{log} \big( \frac{r}{\phi_i^\dagger \phi^i} \big) \ .
\end{eqnarray}
Componentwise, the gauge field, for examples, is determined by
\begin{eqnarray}\label{gauge}
  A_\mu = \frac{1}{2i \phi_i^\dagger \phi^i}
  \Big(\phi_i^\dagger \partial_\mu \phi^i
  - \partial_\mu \phi_i^\dagger \phi^i
  - i \bar \psi_i \bar\s_\mu \psi^i \Big)\ ,
\end{eqnarray}
which implies that above procedure can be understood as supersymmetric
version of solving the Gauss law in GLSM.
Inserting the result back into the Lagrangian, one can finally obtain
\begin{eqnarray}
  \CL^\text{IR} = r \int d^4\th \ \Big[ \text{log} \big(
  \sum_i \phi_i^\dagger \phi^i \big) \Big]\ .
\end{eqnarray}
Assuming one of matter fields, say $\phi^0$, does not vanish,
one can rewrite the above Lagrangian as
\begin{eqnarray}\label{lag1}
  \CL^\text{IR} = r \int d^4\th \ \Big[ \text{log} \big( \phi_0^\dagger \phi^0 \big)
  + \text{log} \big( 1 + Z_m^\dagger Z^m \big) \Big]
  = r \int d^4\th \  \Big[ \text{log} \big( 1 + Z_m^\dagger Z^m \big) \Big] \ ,
\end{eqnarray}
where we used for the last equality the chirality of $\phi^0$. (\ref{lag1})
is precisely the lagrangian for the $\CN=(2,2)$ supersymmetric non-linear
sigma model with target space ${\mathbb CP}^N$. Here
chiral superfields $z^m$ ($m=1,2,...,N$) are defined as
\begin{eqnarray}
  Z^m = \frac{\phi^m}{ \phi^0} \ ,
\end{eqnarray}
from which one can identify it bosonic and fermionic part as
\begin{eqnarray}
  z^m = \frac{\phi^m}{\phi^0}\ , \qquad
  \chi^m = \frac{1}{(\phi^0)^2} \big( \phi^0 \psi^m - \psi^0 \phi^m \big)\ .
\end{eqnarray}

The model we are eventually interested in is a massive version of this
theory. The so-called twisted masses can be introduced by gauging
the flavor symmetry $U(N+1)$
and give expectation values to the
corresponding twisted chiral field $\hat \S$ as
\begin{eqnarray}
  \langle \hat \Sigma \rangle = \text{diag}\big( \langle \hat \S_0 \rangle,
  \langle \hat \S_1 \rangle,..,\langle \hat \S_N \rangle \big) = \begin{pmatrix}
  m_0 & & & \\ & m_1 & & \\ & & \ddots & \\
  & & & m_n \end{pmatrix} \ .
\end{eqnarray}
These vev acts as mass terms for the chiral multiplets, and can be
incorporated into the Lagrangian as
\begin{eqnarray}\label{GLSMlag2}
  \CL = \int d^4\th \ \Big[\phi^\dagger_i e^{-2V } \phi^i e^{2 \langle \hat V_i \rangle }- \frac{1}{4e^2} \bar \Sigma
  \Sigma \Big] - \text{Im}\Big[ \t  \int d^2 \hat \theta \ \S \Big] \ .
\end{eqnarray}
With these twisted masses, there are $N+1$ classical discrete
vacua in this theory. They correspond to
\begin{equation}
 \s = m_i \; ,\;\; |\phi^i|^2=r\;\;{\rm and}\;\; \phi^k=0\;,\;\; k\neq i
\end{equation}
for each $i=0,1,\dots,N$. With such discrete set of vacua, various
topological kink solitons are present, which are the objects of
our interest. One can show that this massive theory flows down to
\begin{eqnarray}\label{massivecpn}
  \CL^\text{IR}_\text{mass}
  = r \int d^4\th \  \Big[ \text{log} \big( 1 + z_m^\dagger
  e^{2 \langle \hat V_m \rangle - 2 \langle \hat V_0 \rangle } z^m \big) \Big] \ .
\end{eqnarray}
In this article, we will be classifying and counting BPS multiplets
of this theory, with a care given to quantum degeneracy and wall-crossing
in weak coupling regime $r \gg 1$ of the sigma model.

The FI parameter $r$ indeed receives the quantum correction at one-loop
level, which leads to the RG running of renormalized FI parameter $r(\mu)$
\begin{eqnarray}
  \mu \frac{\partial}{\partial \mu} r(\mu) = - \frac{N+1}{2\pi}\
  \to \
  r(\mu) \simeq \frac{N+1}{2\pi} \text{log}
  \Big[ \frac{\mu}{\L_\s} \Big]\ ,
\end{eqnarray}
where $\L_\s$ denotes the RG-invariant dynamical scale where
the perturbative analysis breaks down. In order to
rely on our analysis in the article, we therefore have
to introduce sufficiently large twisted masses $m^i$
\begin{eqnarray}
  e \gg |m^i-m^j| \gg \L_\s \nonumber\ ,
\end{eqnarray}
such that the renormalized coupling $r(\mu)$ are
frozen in the weak-coupling regime. On the other hand,
the low-energy theory of (\ref{GLSMlag}) in another
interesting parameter region $e \ll \L$ have been
explored in \cite{Hanany:1997vm,Dorey:1998yh} to study
the BPS states in $\mathbb{CP}^N$ model at strong coupling,
which will be briefly discussed in section 5.
It has been shown that there is the discrepancy between BPS spectra
at weak and strong coupling of the theory, which strongly
implies the existence of curves of marginal stability
somewhere at strong coupling region. Quantum
aspects of central charges and strong/weak coupling marginal
stability walls were also  recently investigated in Ref.~\cite{Shifman:2006bs,Olmez:2007sg}.

As emphasized again,
we will explore the curves of marginal stability and
wall-crossing phenomena not in strong-coupling regime
but in weak-coupling regime.

\paragraph{conserved charges}
For later convenience, we summarize some conserved charges.
The bosonic part of energy functional of this theory
takes the following simple form
\begin{eqnarray}
  \CE = \int dx^3\
  \sum_i \Big[ |D_0 \phi^i|^2 + |D_3 \phi^i|^2 + |\s - m_i|^2 |\phi^i|^2  \Big]\ .
\end{eqnarray}
In the infrared, one can express the energy functional
in terms of sigma model variables as
%
\begin{eqnarray}\label{energy}
  \CE &=& r \int d{\bf x}^3 \ \Big[ \frac{(1+\bar z \cdot z) \d^m_n - \bar z_n z^m}{(1+\bar z \cdot z)^2}
  \big( \dot{\bar z}_m \dot{z}^n + \partial_3 \bar z_m \partial_3 z^n \big) \nonumber\\
  && \hspace*{0.3cm} + \frac{1}{(1+\bar z\cdot z)^2}\sum_n |m_{n}-m_0|^2 |z^n|^2 \nonumber \\
  && \hspace*{0.3cm} + \frac{1}{(1+\bar z\cdot z)^3} \sum_{n<p}
  (m_n-m_p)^2|z_n|^2 |z_p|^2 \big( 1 + |z_n|^2 + |z_p|^2 \big) \nonumber \\
  && \hspace*{0.3cm} + \frac{1}{(1+\bar z\cdot z)^3}
  \sum_{n\neq p\neq q}(m_n-m_p)(m_n-m_q)|z_n|^2|z_p|^2|z_q|^2 \Big]\ .
\end{eqnarray}

Introducing the twisted mass terms, flavor symmetry group $SU(N+1)$ of
$\mathbb{CP}^N$ model is spontaneously broken down to $U(1)^{N}$.
Those charges are defined by following:
$N$ $U(1)$ charges can be parameterized by a following $N+1$-vector
\begin{eqnarray}
  \vec Q = \big( Q_0 , Q_1, .. , Q_{N} \big) \ ,
\end{eqnarray}
where each component is given by
\begin{eqnarray}\label{flavor charge}
  Q_0 &=&  -i\int d{\bf x}^3 \ \phi_0^\dagger D_0 \phi^0 + \text{c.c.}
  \nonumber \\
  &=& r \int d{\bf x}^3 \   \frac{i \sum_m \big(\bar z_m \partial_0 z^m - \partial_0z_m z^m\big)}
  { \big( 1 + \sum_m \bar z_m z^m\big)^2} \ ,
  \\
  Q_n &=&  - i \int d{\bf x}^3 \ \phi_n^\dagger D_0 \phi^n + \text{c.c.}
  \nonumber \\
  &=&  r  \int d{\bf x}^3 \ \frac{- i \big( \bar z_n \partial_0 z^n
  - \partial_0 \bar z_n z^n\big)}{1+\sum_m \bar z_m z^m}
  + \frac{\bar z_n z^n \cdot i \sum_m
  \big( \bar z_m \partial_0 z^m - \partial_0 \bar z_m z^m \big)}
  {\big( 1 + \sum_m \bar z_m z^m\big)^2} \ . \nonumber
\end{eqnarray}
Note that the charge components $Q_i$ ($i=0,1,..N$) always
satisfy the traceless condition
\begin{eqnarray}
  Q_0 + Q_1 + .. + Q_{N} = 0 \ .
\end{eqnarray}

\paragraph{central charge}

Finally let us recall the expression of central charge $Z$ for
$\CN=(2,2)$ massive $\mathbb{CP}^N$ model. Based on the
two-dimensional Witten effect and simple
BPS spectra of (\ref{massivecpn}) such as
fundamental excitations and kink solutions, central charge $Z$
takes the following form at weak coupling limit $r \gg 1$
\begin{eqnarray}\label{central}
  Z = \sum_i  m^i \big( Q_i + \t T_i \big) \ ,
  \qquad
  \t = \frac{\th}{2\pi} - i r \ ,
\end{eqnarray}
as discussed in \cite{Dorey:1998yh}. Here $T$
denotes the topological charge associated with kinks. Because the
theory possesses $N+1$ discrete vacua, $T$ naturally live
in the $SU(N+1)$ root lattice. For a topological kink from vacuum j to
vacuum i, our convention is such that $T_j=-1$, $T_i=1$, and $T_k=0$ for
$k\neq j,i$.

The exact expression for central charge $Z$ has also proposed
in \cite{Hanany:1997vm} as
\begin{eqnarray}\label{central1}
  Z = \sum_i \big( m^i Q_i  + m_D^i T_i \big) \ ,
  \qquad
  m_D^i = \CW(e_i)\ ,
\end{eqnarray}
where $e_i$ are determined by roots of the
polynomial equation
\begin{eqnarray}\label{central2}
 \prod_i \big( x - m_i \big) - \L^{N+1}_\s =
 \prod_i \big( x - e_i \big) = 0 \ ,
\end{eqnarray}
and $\CW(e_i)$ are given by
\begin{eqnarray}\label{central3}
 \CW(e_i) = \frac{N+1}{2\pi} e_i + \sum_i \frac{m_i}{2\pi}
 \text{log}\Big[ \frac{e_i - m_i}{\mu} \Big]\ .
\end{eqnarray}
We will discuss in Section 5 an interesting implication of the
exact expression of central charge $Z$ in relation to
four-dimensional $\CN=2$ supersymmetric gauge theories.


\subsection{BPS equations}

The supersymmetry transformation for $z^m$ can be read off from
those of GLSM fields: for examples, the variation rules for
fermions $\chi^m$ are given by
\begin{eqnarray}\label{CPNtransform}
  \d \chi^m = \frac{1}{(\phi^0)^2} \big( \phi^0 \d \psi^m
  - \d \psi^0 \phi^m \big) + \cdots\ ,
\end{eqnarray}
where we suppressed the irrelevant terms in our discussion.
The transformation rules for GLSM fermion fields $\psi^i$
are given by
\begin{eqnarray}\label{GLSMtranform}
  \d \psi^i = \t^3 \e D_3 \phi^i + \e D_0 \phi^i - i \t^I \e
  \big( \s_I \phi^i - \phi^i m_I^i\big)\ ,
\end{eqnarray}
where $I$ run form $1,2$. Here we  substitute
(\ref{gauge}) for the GLSM gauge fields:
\begin{eqnarray}\label{gauge2}
  A_\mu = \frac{\bar z_m  \partial_\mu z^m - \partial_\mu \bar z_m \cdot z^m}
  {2i \big(1 + \bar z_m z^m \big)} + \cdots\ ,
\end{eqnarray}
and also substitute the following for the GLSM vector scalar $\s$
\begin{eqnarray}\label{sigma}
  \s = \frac{m_0 + m_n \bar z_n z^n}{ 1+ \bar z_m z^m} + \cdots\ ,
\end{eqnarray}
with $m_i \equiv (m^i)_1 - i (m^i)_2$. We dropped again
the fermion contribution here, which are
irrelevant in our discussion below.

Inserting the above results (\ref{GLSMtranform}) back into (\ref{CPNtransform}),
BPS solitons of $\mathbb{CP}^N$-model should satisfy the following condition
\begin{eqnarray}\label{BPSeqn}
  \phi^0 \big( \t^3 \e D_3 \phi^n + \e D_0 \phi^n
  + i \hat \t_{m_n} \e \phi^n \big) - \phi^n \big(
  \t^3 \e D_3 \phi^0 + \e D_0 \phi^0 \big) = 0\ .
\end{eqnarray}
where $\hat \t_{m_n}$ is defined as
\begin{eqnarray}
  \hat \t_{m_n} \equiv \tau^I (m^n - m^0)_I
  = \begin{pmatrix} & m_{n0} \\ \bar m_{n0} & \end{pmatrix}\ ,
  \qquad m_{n0} = m_{n} - m_0\ .
\end{eqnarray}

\subsection{BPS (multi-)kinks}

\paragraph{simple BPS kinks}

Let us first review BPS kinks solutions.
Since they are static particle, the BPS equation (\ref{BPSeqn})
can be simplifies as
\begin{eqnarray}
  \t^3 \big( D_3 \phi^n - z^n D_3 \phi^0 +
  i \t^3 \hat \t_{m_n} \phi^n \big) \e = 0 \ .
\end{eqnarray}
As referred to appendix for detailed computation, one can show that
\begin{eqnarray}
  D_3 \phi^n - z^n D_3 \phi^0 = r \frac{\partial_3 z^n}{\sqrt{1+\bar z_n z^n}}\ ,
\end{eqnarray}
from which one can massage the above BPS equation into
\begin{eqnarray}
 \Big[ \frac{\partial_3 z^n}{\sqrt{1+\bar z_n z^n}} + i \t^3 \hat \t_{m_n}
 \frac{z^n}{\sqrt{1+\bar z_n z^n}}\Big] \e = 0\ .
\end{eqnarray}
Since $\big(\t^3 \hat \t_{m_n}\big)^2=-|m_{n0}|^2$, the BPS equation is
finally given by
\begin{eqnarray}
  \partial_3 z^n \pm |m_{n0}| z^n = 0 \ , \qquad
  z^m = 0 \ \ \text{ for } m\neq n\ ,
\end{eqnarray}
provided that $m_n \neq m_n$. The solutions are
therefore given by
\begin{eqnarray}
  z^n = \text{exp}\Big[ \pm |m_{n0}| ({\bf x}^3-{\bf x}_0) \Big]\, .
\end{eqnarray}
The energy of this configuration saturate a topological energy bound
since
 \begin{eqnarray}
    \CE &=& r \int d{\bf x}^3 \ \Big[ \frac{1}{(1+\bar z_n  z^n)^2}
    \big|  \partial_3 z^n \mp |m_{n0}| z^n \big|^2
    \pm \frac{|m_{n0}|}{(1+\bar z  z)^2} \partial_3 \big(\bar z_n z_n\big) \Big]
    \nonumber \\
    &\geq& - r |m_{n0}| \Big[ \frac{1}{1+\bar z_n  z^n}
    \Big]^{{\bf x}^3=+\infty}_{{\bf x}^3=-\infty} = r |m_{n0}|\,.
  \end{eqnarray}

\paragraph{composite  kinks}

Let us denote a BPS kink which interpolates
from $m$th vacuum to $n$th vacuum as $nm$-kink.
Suppose that the phases of two mass-parameters $m_{10}$ and
$m_{20}$ are aligned as parallel. Without loss of generality,
one can set $|m_{20}| > |m_{10}|$.
Then, the $20$-kink can be also understood as a bound state
of a $10$-kink and a $21$-kink:
the BPS equations for $20$-kink are
\begin{eqnarray}
  \partial_3 z^1 + i \t^3 \hat \t_{m_{10}} z^1 = 0 \ , \qquad
  \partial_3 z^1 + i \t^3 \hat \t_{m_{20}} z^1 = 0\ , \qquad
  \big[ \t^3 \hat\t_{m_{10}} , \t^3 \hat\t_{m_{20}} \big] = 0 \ ,
\end{eqnarray}
or equivalently
\begin{eqnarray}
  \partial_3 z^1 \mp |m_{10}| z^1 = 0\ , \qquad
  \partial_3 z^2 \mp |m_{20}| z^2 = 0\ .
\end{eqnarray}
The solution then turns out to be
\begin{eqnarray}
  z^1 =  \text{exp}\Big[ \pm |m_{10}| {\bf x}^3 \Big]\ ,
  \qquad
  z^2 =  \text{exp}\Big[ \pm |m_{20}| ({\bf x}^3 -{\bf x}_0)\Big]\ ,
\end{eqnarray}
after a suitable choice of the origin. Here ${\bf x}_0$ parameterizes
the relative distance between constituent BPS kinks. Note that
the phase factor of each $z^m$ describes one-parameter degeneracy
of such kink solutions, so in fact we can have an arbitrary complex
number multiplying each of $z^{1,2}$'s. The fact that they have
the same energy can be directly checked. See Appendix A.

Obviously, this can be repeated for other $z_m$'s straightforwardly.
When all $m_{n0}$'s are aligned in the complex plane, the general
solution is
\begin{eqnarray}
  z^m =  \zeta^m\text{exp}\Big[ \pm |m_{m0}| {\bf x}^3 \Big]\
\end{eqnarray}
with arbitrary complex numbers $\zeta^m$'s which are moduli coordinates
of the soliton.

\begin{figure}
\begin{center}
\includegraphics[width=8cm]{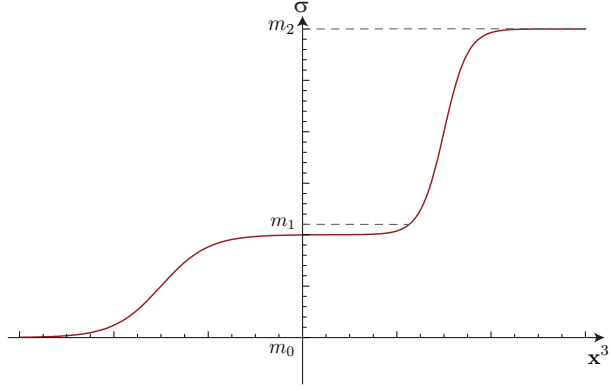}
\caption{Configuration of the GLSM field $\sigma$.
It implies that the system is placed in $\sigma=m_1$
vacuum at ${\bf x}^3 = - \infty$, and in $\sigma=m_2$
vacuum at ${\bf x}^3 = + \infty$. The size of the plateau
near $m_1$ is determined by how far 10-kink and 21-kink are
separated, which is in turn determined by certain ratio
between $\zeta^1$ and $\zeta^2$.}\label{sigmaprof}
\end{center}
\end{figure}
The GLSM $\sigma$ field (\ref{sigma}) is useful
for describing the general behavior of the kink solution,
which is depicted in Figure \ref{sigmaprof} for this solution.
For a finite ${\bf x}_0$,
$\s$ starts with the vacuum $\s=m_0$,
approaches the vacuum $\s=m_1$ (never touches it),
and eventually goes to the vacuum $\s=m_2$ as ${\bf x}^3$
increases. This shows that the solution indeed a sequential
sum of 10-kink and 21-kink.


\subsection{zero modes}

Here we briefly dwell on details of
fermion zero mode counting. Bosonic ones were already noted in
previous section: there is one complex bosonic collective
coordinate for each $z^n$ kink, provided that all masses $m_{n0}$ are
of the same phase. We will find below that for each $z_n$ kink
there is also one complex fermionic zero modes.
The linearized fermion equations of motion are given by
\begin{eqnarray}
  \bar \s^M \big( D_M \chi^n + D_M z^m \G^n_{\ ml} \chi^l \big) = 0 \ ,
\end{eqnarray}
with
\begin{eqnarray}
  \G^n_{ml} = - \frac{\d^n_l \bar z_m + \d^n_m \bar z_l}{1+\bar z \cdot z}
  \ , \qquad
  \G^\bn_{\bm \bl} = - \frac{\d^\bn_\bl z_\bm
  + \d^\bn_\bm z_\bl}{1+\bar z \cdot z}\ ,
\end{eqnarray}
where the covariant derivatives are defined as
\begin{eqnarray}
  D_M \chi^n = \partial_M \chi^n + i \big( \hat{A}^n_M - \hat{A}^0_M \big)
  \chi^n \ .
\end{eqnarray}
Here $M$ run from $0,1,2,3$. The twisted mass terms are written as if
it is gauge field along $2,3$ directions, and contributes
\begin{eqnarray}
  \bar \s^M \big( \hat{A}^n_M - \hat{A}^0_M \big) =
  -\hat \t_{m_{n0}} = - \begin{pmatrix}
  0 & m_n- m_0 \\ \bar m_n -\bar m_0 & 0
  \end{pmatrix}\ .
\end{eqnarray}
Clearly the derivative $\partial_M$ runs only for $M=0,1$.

For simplicity let us again take the example of a double-kink
with aligned masses $|m_{20}|>|m_{10}|>0$.
The BPS solution in this case was
\begin{eqnarray}
&&  z^1 =  \zeta^1\text{exp}\Big[  |m_{10}| {\bf x}^3 \Big]\, ,\nonumber\\
 &&   z^2 =  \zeta^2\text{exp}\Big[  |m_{20}| {\bf x}^3 \Big] \,.\nonumber
\end{eqnarray}
Recall that, despite its deceptively simple appearance, the solution should
be viewed as a combination of two kinks, one from 0 to 1 and another
from 1 to 2, which will interact with each other when one begins to
move them around. The fermionic zero modes
in this background are equally simple and deceptive. There are exactly one zero
mode for each $\chi$, and we find (in the limit of $\zeta^1=0$)
\begin{eqnarray}
&&  \chi^{1}_0 = e^{|m_{10}|{\bf x}^3}  \e_0 \nonumber\\
&&  \chi^{2}_0 =  e^{|m_{20}|{\bf x}^3} \e_0 \ .
\end{eqnarray}
with the constant spinor obeying $i\t^3\hat \t_{m_{20}} \e_0 = - |m_{20}|\e_0 $.

The Goldstino mode, in the limit $|\zeta^1|\ll1$,
is the combination $\chi^{1,2}=\zeta^{1,2}e^{|m_{10,20}|{\bf x}^3} \e_0$,
quantization of which endows the soliton with the basic BPS multiplet structure.
The other combination is more interesting. This is
a superpartner to the nontrivial bosonic moduli of the kinks that encodes
relative separation and mutual interaction of 10-kink and 21-kink.
We will come back to them later when we search for quantum spectrum of
flavored kinks.

\section{Flavored kink solitons and marginal stability}


Since the theory has $U(1)^N$
flavor charges, BPS objects may carry both topological
and flavor charges. A kink with generic flavor charge will be
called flavored kinks. We present in this section the explicit
construction of flavored kink solitons together with
preliminary discussion on their marginal stability behavior.
An important fact here is that these generic flavored kinks
appears only when the mass parameters of the theory is misaligned,
i.e., when they are no longer lined up in the complex plain.
This is analogous to (dis-)appearance of generic dyons in $D=4$ $\CN=2$
SYM and also of 1/4 BPS dyons in $D=4$ $\CN=4$ SYM, depending on
how the vacuum expectation values of adjoint scalar fields are
aligned or misaligned.

In order to investigate the dyonic spectrum of
the two-dimensional $\mathbb{CP}^{N}$ model, let us
introduce the time-dependence on the phase factor of
sigma model fields $z^m$. Then, the BPS equation
(\ref{BPSeqn}) can be rewritten as
\begin{eqnarray}
  \t^3 \big( D_3 \phi^n - z^n D_3 \phi^0 +
  i \t^3 \hat \t_{m_n} \phi^n \big) \e
  + \big( D_0 \phi^0 - z^n D_0 \phi^n \big) \e = 0 \ .
\end{eqnarray}
Inserting (\ref{gauge2}) into the above equation,
one can show that flavored kinks should satisfy the following
\begin{eqnarray}\label{timeBPSeqn}
 \Big[ \tau^3 \frac{\partial_3 z^n}{\sqrt{1+\sum_m\bar z_m z^m}} + i  \hat \t_{m_n}
 \frac{z^n}{\sqrt{1+\sum_m \bar z_m z^m}} +
 \frac{\partial_0 z^n}{\sqrt{1+\sum_m \bar z_m z^m}}  \Big] \e = 0\ .
\end{eqnarray}

\subsection{simple flavored kinks}

Let us again review  simple flavored kink solutions whose
topological charge and flavor charge are parallel \cite{Dorey:1998yh}.
In this case, without loss of generality, one can turn off
all complex field $z^n$ expect one, say $z^1$.

Then, the above BPS equations (\ref{timeBPSeqn}) can be simplified as
\begin{eqnarray}
  \big( \partial_0 z^1 + i \hat \t_\text{E} \big) \e +
  \t^3 \big( \partial_3 z^1 + i \t^3 \hat \t_\text{M} \big) \e = 0\ ,
\end{eqnarray}
where $\hat\t_\text{E,M}$ are defined by
\begin{eqnarray}
  \hat \t_\text{E} + \hat \t_\text{M} = \hat \t_{m_{10}}\ .
\end{eqnarray}

In order to have solutions to this equation,
we have to demand the projectors $\hat\t_\text{E,M}$
to satisfy the following compatibility condition
\begin{eqnarray}
  \big[ \hat\t_\text{E} , \t^3 \hat \t_\text{M} \big] = 0\ .
\end{eqnarray}
\begin{figure}[t]
  \begin{center}
  \includegraphics[width=12cm]{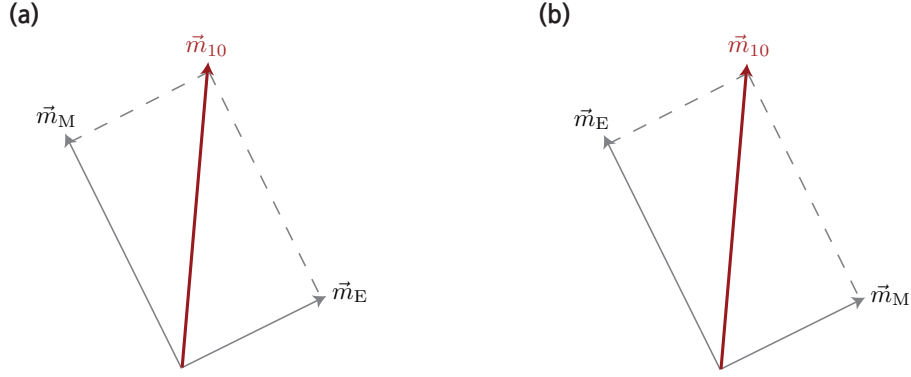}
  \caption{For a simple flavor kink, the mass parameter $\vec m_{10}$
  can be decomposed into arbitrary two orthogonal vectors
  ${\vec m}_\text{M}$ and ${\vec m}_\text{E}$.
  For (a), $m_\text{M}$ lies on the right hand side of $m_{10}$ while for
  (b) $m_\text{M}$ lies on the left hand side of $m_{10}$.}
  \end{center}\label{charge1}
\end{figure}
One can easily find a family of solution, parameterized by
\begin{eqnarray}
  \hat \t_\text{E} = \vec \t \cdot {\vec m}_\text{E} \ ,
  \qquad
  \hat \t_\text{M} = \vec \t \cdot {\vec m}_\text{M} \ ,
\end{eqnarray}
where vectors $\vec{m}_\text{E}$ and $\vec{m}_\text{M}$
are orthogonal decomposition of $\vec m_{10}$ as depicted in
figure \ref{charge1}.

For the case (a), 
the flavored kink solution is 
\begin{eqnarray}\label{simpleflavorkink}
  z^1 = \text{exp}\Big[ \pm |m_\text{M}| {\bf x}^3 \pm i |m_\text{E}|t \Big]\ ,
\end{eqnarray}
%
For the case (b), the flavor kink solution is instead given by
\begin{eqnarray}
  z^1 = \text{exp}\Big[ \pm |m_\text{M}| {\bf x}^3 \mp i |m_\text{E}|t \Big]\ .
\end{eqnarray}
Without loss of generality, let us concentrate on
the case (a). Some conserved charges of the simple
flavored kink solutions are in order.

\paragraph{flavor charge}

For a simple flavored kink, the nonvanishing flavor charges (\ref{flavor charge}) are
\begin{eqnarray}
  Q_1 = -Q_0= \pm r \int_{-\infty}^{+\infty} d{\bf x}^3  \ \frac{|m_\text{E}|}
  {2 \text{cosh}^2(|m_\text{M}| {\bf x}^3)} = r \frac{|m_\text{E}|}{|m_\text{M}|}\ .
\end{eqnarray}
%

\paragraph{energy}

For the simple flavored kinks, the energy functional (\ref{energy}) can be
massaged into a sum of complete squares like
\begin{eqnarray}
  \CE &=& r \int d{\bf x}^3 \  \frac{1}{(1+ |z^1|^2 )^2}
  \bigg[
  \big|  \partial_3 z^1 \mp |m_\text{M}| z^1 \big|^2
  + \big| \partial_0 z^1 \mp i |m_\text{E}| z^1 \big|^2
  \nonumber \\ && \hspace*{4.5cm}
  \mp |m_\text{E}| i \big( \bar z_1 \partial_0 z^1 - \partial_0 \bar z_1 z^1 \big)
  \pm |m_\text{M}| \partial_3 \big(\bar z_n z_n\big) \bigg]
  \nonumber \\
  &\geq& \mp |m_\text{E}| Q_0
  \mp r |m_\text{M}| \left. \frac{1}{1+\bar z_n  z^n}
  \right|^{{\bf x}^3=+\infty}_{{\bf x}^3=-\infty}
  = \pm \frac{r |m_{10}|^2}{|m_\text{M}|}\ ,
\end{eqnarray}
where we used  $|m_\text{M}|^2 + |m_\text{E}|^2 = |m_{10}|^2$.
Since
\begin{eqnarray}
  Z = - m_{10} Q_0 + i r m_{10} = r \frac{|m_{10}|}{|m_\text{M}|} e^{i\varphi_{m_{10}}}
  \big( - |m_\text{E}| + i |m_\text{M}| \big)=
   \frac{r |m_{10}|^2}{|m_\text{M}|} e^{i \varphi_{m_\text{E}}}\ ,
\end{eqnarray}
the solutions are indeed BPS with $ \CE = |Z|\ .$

\subsection{composite flavored kinks and marginal stability}

It has been noted previously that the solitonic sector
of this $D=2$ QED has some features reminiscent of certain $D=4$
Seiberg-Witten theory, where the topological charge and the
flavor charges are mapped to the magnetic charge and the electric
charges, respectively.  On the other hand, dyonic solitons in
the $\CN=2$  supersymmetric gauge theories in four dimensions
are such that magnetic and electric charges are generically not parallel
\cite{Lee:1998nv,Bergman:1997yw}. This is in turn related to
existence of multi-pronged strings in string theory.

These class of $D=4$ BPS states are useful in that
one can study the issue of marginal
stability in weakly-coupled regime of the theory.
In this subsection, we will look for their analog in $D=2$ theory,
considering  flavored kinks whose topological and flavor charge
are not parallel misaligned,  and discuss
their marginal stability briefly. In section 4, their
quantum spectrum and wall-crossing phenomena
will be explored in more details.

\begin{figure}[t]
\begin{center}
\includegraphics[width=12cm]{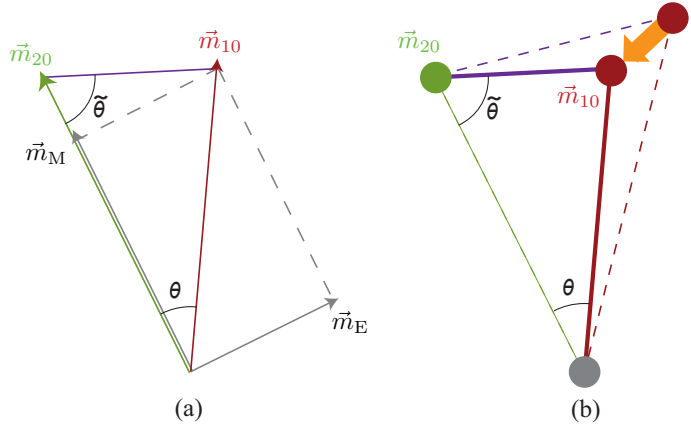}
\caption{(a) Schematic diagram for decomposition of
the mass parameter $\vec m_{10}$. Let us denote
the relative angle between two mass parameters
$m_{10}$ and $m_{20}$ by $\th$.  By definition, $\vec m_{\text M} $ is
parallel to $\vec m_{20}$.
We are considering cases where $|\vec m_{\text M}|<| m_{20}|$. (b)
Each node denotes
the vacuum of the theory, i.e., grey for $\s=m_0$,
red for $\s=m_1$ and green for $\s=m_2$. The solid
lines schematically describe the GLSM $\s$ field. It
somehow parallels with the four-dimensional picture of
pronged strings where each node represents
the D3-brane and solid line denotes
the (p,q)-string. In section 5, the parallel between
$D=2$ sigma models and $D=4$ gauge theories will be
discussed in more details.}\label{nontrivial}
\end{center}
\end{figure}

For simplicity, let us first assume that
\begin{eqnarray}
  z^1 = z^1({\bf x}^3,t) \ , \qquad
  z^2 = z^2({\bf x}^3)\ , \qquad
  z^m = 0 \ \ \text{ for } m\neq 1,2\ .
\end{eqnarray}
For this ansatz, the BPS equation (\ref{timeBPSeqn}) can be
rewritten as
\begin{eqnarray}
  \Big[ \partial_3 z^2 + i \t^3 \hat \t_{m_{20}} z^2 \Big]\e &=& 0 \ ,
  \nonumber \\
  \Big[\t^3 \partial_3 z^1 + \partial_0 z^1 + i \hat \t_{m_{10}} z^1 \Big]\e &=& 0 \ .
\end{eqnarray}
Guided by the previous example of simple flavored kink,
let us rewrite the second equation into the following form
\begin{eqnarray}
   \t^3\Big[ \partial_3 z^1 + i \t^3 \hat \t_{m_\text{M}}z^1 \Big]\e
   + \Big[\partial_0 z^1 + i \hat \t_{m_\text{E}} z^1\Big] \e = 0
   \ ,   \qquad \hat \t_\text{E} + \hat \t_\text{M} = \hat \t_{m_{10}}\ .
\end{eqnarray}
In order to find out half-BPS solutions,
we therefore have to demand three projectors to commute
to each other
\begin{eqnarray}\label{condition1}
  \big[\t^3 \hat \t_{m_{20}} , \t^3 \hat \t_{m_\text{M}} \big]=0\ , \qquad
  \big[\t^3 \hat \t_{m_\text{M}} , \hat \t_{m_\text{E}} \big] = 0 \ , \qquad
  \big[\t^3 \hat \t_{m_{20}}, \hat \t_{m_\text{E}} \big] = 0 \ .
\end{eqnarray}
One can again easily parameterize the solutions of the above
relations as
\begin{eqnarray}
  \hat \t_\text{E} = \vec \t \cdot {\vec m}_\text{E} \ ,
  \qquad
  \hat \t_\text{M} = \vec \t \cdot {\vec m}_\text{M} \ ,
\end{eqnarray}
where vectors $\vec{m}_\text{E}$ and $\vec{m}_\text{M}$ are depicted in
figure \ref{nontrivial}.
The BPS solutions of interests are
\begin{eqnarray}
  z^1 = \text{exp}\Big[ |m_\text{M}| {\bf x}^3 + i |m_\text{E}| t \Big]
  \ , \qquad
  z^2 = \text{exp}\Big[ |m_{20}| ({\bf x}^3 - {\bf x}_0 )\Big] \ ,
\end{eqnarray}
after a suitable choice of origin of ${\bf x}^3$.


\paragraph{flavor charge and marginal stability}

For the above  solution, the flavor charges (\ref{flavor charge})
are
\begin{eqnarray}
  Q_0 &=&  -2 r |m_\text{E}| \ \int_{-\infty}^{+\infty} d{\bf x}^3 \
  \frac{|z^1|^2}{\big(1+ |z^1|^2 + |z^2|^2\big)^2}\ , \nonumber \\
  Q_2 &=&  -2 r |m_\text{E}| \ \int_{-\infty}^{+\infty} d{\bf x}^3 \
  \frac{|z^1|^2 |z^2|^2}{\big(1+ |z^1|^2 + |z^2|^2\big)^2} \ , \nonumber \\
  Q_1 &=& - Q_0 - Q_2 \ .
\end{eqnarray}
When we place the mass parameter $m_{10}$ on
a so-called wall of marginal stability, as depicted in figure \ref{nontrivial} (b),
the relative distance ${\bf x}_0$ diverges such that
the $20$-flavored kink decays into two constituent $10$- and $21$-flavored
kinks. This is an underlying physical reason for the phenomenon
of wall-crossing.  At wall-crossing, one can easily show that
the GLSM field $\s$ actually turn touches the vacuum $\s=m_1$,
as described in figure \ref{nontrivial} (b).

For classical soliton whose flavored charges are not quantized,
this can be viewed backward as a process where the flavor charges are
increased until the kink solution decompose into two. This
``maximal'' or ``critical'' flavor charge can
can be read off from the solution as
\begin{eqnarray}
  Q_0^\text{cr} \simeq
  - r \tan \th \ , \qquad
  Q_2^\text{cr} \simeq
  - r\tan \tilde \th\ , \qquad
  Q_1^\text{cr} \simeq + r \big( \tan \th + \tan \tilde \th \big)\ .
\end{eqnarray}
With quantized (and thus fixed) flavor charges, we can use this formula to determine
the critical values of $\theta $ and $\tilde\theta$, which in turn
determine the marginal stability wall for breaking this soliton to
a simple flavored 10-kink and a simple flavored 21-kink.

%

\paragraph{central charge}

As discussed before,
the central charge of the present model can take the following form
\begin{eqnarray}
  Z = \sum_n  m^n \big( Q_n + \t T_n \big) \ ,
  \qquad
  \t = \frac{\th}{2\pi} - i r \ .
\end{eqnarray}
For the composite flavored kinks, the central charge $Z_{20}$
can be decomposed into those of constituent particles, say
\begin{eqnarray}
  Z_{20} = Z_{10} + Z_{21}\ ,
  \qquad
  Z_{10} = - m_{10} Q_0 + \t m_{10}\ , \
  Z_{21} = + m_{21} Q_2 + \t m_{21}\ .
\end{eqnarray}
On the wall of marginal stability where the flavor charges take
their critical values $\vec Q^\text{cr}$,
the central charges of constituent particles become
\begin{eqnarray}
  Z_{10} &=& m_{10}\big( + \tan \th - i \big) \ , \nonumber \\
  Z_{21} &=& m_{21}\big( - \tan{\tilde\th} - i \big)\ .
\end{eqnarray}
Note that, on the wall of marginal stability,
the phases of two mass-parameters satisfy the relations below
\begin{eqnarray}
  \th + \tilde \th = \varphi_{m_{21}} - \varphi_{m_{10}}\
  \nonumber,
\end{eqnarray}
from which one can conclude that phase difference between
$Z_{10}$ and $Z_{21}$ is
\begin{eqnarray}
  \text{arg}\big(Z_{21}\big)  - \text{arg}\big(Z_{10}\big) =
  - \tilde \th - \th + \varphi_{m_{21}}  -\varphi_{m_{10}} = 0 \ !
\end{eqnarray}
As expected, we find that phases of the two central charges $Z_{10},Z_{21}$
coincides at the marginal stability wall.

\section{Quantum BPS states and wall-crossing}

\subsection{low energy interactions of kinks}

In this section, we construct and count quantum BPS
states of topological kinks with flavor charges, by
studying the low energy interactions of simple kinks.
When $m_{i0}$ are all of same phase, each kink carries
one complex bosonic moduli, and their moduli space
is naturally K\"ahler. The holomorphic coordinates $\z^i$'s
are defined in terms of the soliton solution as
\begin{eqnarray}
  z^i = e^{m_{i0}{\bf x}^3} \cdot e^{m_{i0}{\bf x}^i+i\th^i}
  \equiv e^{m_{i0}{\bf x}^3} \z^i \ .
\end{eqnarray}
The moduli space dynamics is obtained by taking
time-dependence of the form $\z^i (t)$ with small velocity
as usual. The K\"ahler potential is found by integrating
the field theory kinetic term as \cite{Tong:2002hi}
\begin{eqnarray}
  K\big(\bar \z, \z\big) =
  \int d{\bf x}^3 \
  \CK \big( \bar z , z \big) \ ,
  = r \int d{\bf x}^3 \
  \text{log}\bigg[ 1+ \sum_i e^{2m_{i0}{\bf x}^3} \bar \z_i \z^i\bigg]\ ,
\end{eqnarray}
from which the moduli space metric follows
\begin{eqnarray}\label{modulimetric}
  g_{i \bj}\big(\z^i,\bar \z_i\big) = r \int d{\bf x}^3 \
  \Bigg[ \frac{e^{2m_{i0}{\bf x}^3}\d_i^j}
  {1+ \sum_k e^{2m_{k0}{\bf x}^3} \bar \z_k \z^k} -
  \frac{e^{2(m_{i0} + m_{j0}) {\bf x}^3 } \bar \z_i \z^j}
  {\big(1+ \sum_k e^{2m_{k0}{\bf x}^3} \bar \z_k \z^k\big)^2}
  \Bigg] \,.
\end{eqnarray}
Here let us first concentrate on $\mathbb{CP}^2$ model, from which
we can read off the indices of all BPS states following an
argument of type found in Ref.~\cite{Stern:2000ie}.

For the moment, let us further assume $m_{20}=2m_{10}$. This
causes two different restrictions on the mass parameters for
our purpose. One is the special ratio between the two absolute
values, which is harmless in counting supersymmetric states.
The other, namely alignment of the two phases, pose a physical
restriction to the spectrum. We will shortly abandon the latter.

The moduli space metric is then compactly written as
  \begin{eqnarray}
    &&g =
    g_\text{com} + g_\text{rel} \ ,
    \qquad
    g_\text{com} = \frac{r}{4m} \Big|d \text{log} \z^2 \Big|^2\ ,
    \quad
    g_\text{rel} = \frac{r}{4m}F({|\z^1|^4/|\z^2|^2}) \Bigg|d \frac{\z^2}{{\z^1}^2}\Bigg|^2\ ,
    \nonumber
  \end{eqnarray}
with
\begin{eqnarray}\label{F1}
F(1/w)=\frac{1}{w(1-4w)} + \frac{2}{(1- 4w)^{3/2}}
    \text{log} \left(\frac{1 - \sqrt{1 -4w}}{1 + \sqrt{1 - 4 w}}\right)\,,
    \end{eqnarray}
    for $ 4w<1 $ and
    \begin{eqnarray}\label{F2}
F(1/w)=  - \frac{1}{w(4w-1)}
    + \frac{4}{(4w-1)^{3/2}}\text{tan}^{-1}\left(\sqrt{4w - 1}\right)\,,
\end{eqnarray}
for $4w>1$. This shows that $\zeta^2$ plays the role of the center of mass
coordinates, while
$$\zeta_{rel}\equiv \zeta^1/\sqrt{\zeta^2}$$
plays the role of the relative coordinate.
It is important  for a later purpose to note that in the limit of $|\z_{rel}|\to \infty$
$g_\text{rel}$ is reduced simply to
  \begin{eqnarray}
    g_\text{rel} \simeq \frac{r}{m}   \biggr|d\z_{rel}/{\z_{rel}}\bigg|^2\ .
  \end{eqnarray}
On the other hand, in the limit of $\z_{rel}
\to 0$,   we have
  \begin{eqnarray}
    g_\text{rel} \sim
    \big| d \z_{rel}\big|^2\ .
  \end{eqnarray}
so $\zeta_{rel}$ is itself a good coordinate near origin where the two kinks coincides in real space.

The phases $\th^{1,2}$ of $\zeta^{1,2}$ are each $2\pi$-periodic and turning on
their (integral) momenta corresponds to turning on $U(1)$ flavor charges of type
$q^{i0}=q^i-q^0$; $q^i$ is the charge of $i$-th diagonal unbroken favor group.
Defining the phase of $\zeta_{cm}$ as $\theta_{cm}$ and $\zeta_{rel}$ as $\varphi$, we find
\begin{equation}
\th_{cm}=\th^2,\qquad \varphi=\th^1-\frac{\th^2}{2} \,,
\end{equation}
and thus
\begin{equation}
q^{10}=q, \qquad q^{20}=q_{cm}-\frac{q}{2}\,,
\end{equation}
where $q_{cm}$ and $q$ are conjugate momenta of $\th_{cm}$ and $\varphi$.
The actual flavor charge for these are
\begin{equation}
(q^0,q^1,q^2,\dots)=(q_{cm}-q/2, q, -q_{cm}-q/2,0,0,\dots) \,.
\end{equation}
Note that  $q$ is integral while $q_{cm}$ should be integral or
half-integral depending on whether $q$ is even or odd.
Such a correlation between
relative and center of mass charges is common, and here due to the identification
\begin{equation}\label{z2}
(\th_{cm},\varphi)\sim (\th_{cm}+2\pi,\varphi-\pi)\,.
\end{equation}
The total moduli space has the form
\begin{equation}
{\mathbb R}\times \frac{[0,4\pi]\times {\cal M}_2}{{\mathbb Z_2}}\,,
\end{equation}
where the relative moduli space ${\cal M}_2$ has a topology of $R^2$
and where ${\mathbb Z}_2$ acts as (\ref{z2}). The center of mass phase
and the quotient action depends on the masses of individual
kinks, in general.

Such a charge state, say with $q_{cm}=0$, precisely corresponds to
the classical solution we find in the previous section with $q=Q_E$.
As we saw there, however, a flavored kink states of this kind do not
appear unless some of the twisted masses are misaligned in the
complex plane. On the other hand, with such misaligned masses,
the composite kink for which we obtained the moduli dynamics is no
longer a solution to the equation of motion unless $\zeta_{rel}=0$.
With $m_{20}=2m_{M}>0 $ and $m_{10}=m_\text{M} + i m_\text{E}$,
the relative moduli space makes sense only if $m_E=0$ while the
flavored kinks appears only if $m_E\neq 0$.

These two issues are in fact tied together. Whenever
$m_{E}\neq 0$, unflavored 20-kink configuration costs
more energy than the central charge bound and this extra energy,
\begin{eqnarray}
  \D \CE = r |m_\text{E}|^2 \int d{\bf x}^3\ \frac{|z^1|^2 \big( 1+ |z^2|^2 \big)}
  {\big( 1+ |z^1|^2 + |z^2|^2\big)^2} \,\;,
\end{eqnarray}
should be interpreted as a potential in the two-kink moduli space
dynamics.

With $m_{20}=2m_\text{M}\equiv2m$, we find
\begin{eqnarray}\label{dyonenergy}
  \D \CE =
  \frac{r |m_\text{E}|^2}{m}
  \frac{|\z^2|^2}{|\z^1|^4} F(|\z^1|^4/|z^2|^2)=\frac{m_E^2}{2}
  g_{rel}\left(\frac{\partial}{\partial\varphi},\frac{\partial}{\partial\varphi}\right)
\end{eqnarray}
Thus, the bosonic part of relative moduli space dynamics must be
modified to
\begin{equation}
L_{rel}=\frac12 (g_{rel})_{\mu\nu}\dot y^\mu \dot y^\nu-\frac12 m_E^2(g_{rel})_{\mu\nu}K^\mu K^\nu
\end{equation}
where
\begin{equation}
K= \frac{\partial}{\partial\varphi}\ .
\end{equation}
happens to be a holomorphic Killing vector field on the moduli space.

\begin{figure}[t]
\begin{center}
\includegraphics[width=8cm]{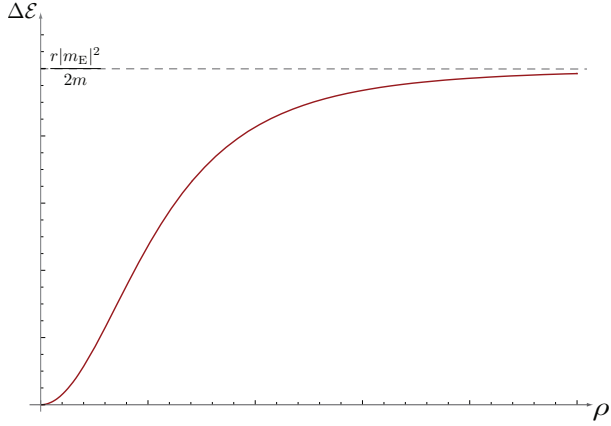}
\caption{The profiles of attractive scalar potential
in the moduli space dynamics of two-kinks system,
induced by tension of composite kinks. }\label{figuredyon}
\end{center}
\end{figure}
This potential energy on the moduli space $\D \CE$,
depicted in figure \ref{figuredyon}, shows physical separations between simple kinks are
no longer a moduli degree of freedom, since it generates an attractive force between the
two kinks. On the other hand, when the conjugate momentum $q$ of $\varphi$ is turned on,
this induces a repulsive angular momentum barrier between the two kinks. For finite
relative charge $q$, then, one can generically expect flavored two kinks states with
the relative position determined by the balance of these two forces. The amount of
the flavor charge, the mass parameter $m_E$, and the size of $\zeta_{rel}$ are all
interrelated, which was shown implicitly in the classical analysis of section 3.

More generally, we may consider $L0$-kink dynamics, regarded as a collection
of 10-kink, 21-kink,32-kink etc, with
$$ m_{p0}=m^{(p)}_M+im^{(p)}_E$$
for $p=1,2,...,L-1$ and $0<m^{(1)}_M < m^{(2)}_M <\cdots< m^{(L-1)}_M  <m_{L0}$.
The above Lagrangian generalizes to
\begin{equation}
L_{rel}=\frac12 (g_{rel})_{\mu\nu}\dot y^\mu \dot y^\nu-\frac12(g_{rel})_{\mu\nu}
 (m_E^{(p)}K_p^\mu) (m_E^{(q)}K_q^\nu) \,,
\end{equation}
where$ K_p$'s are linear combinations of
holomorphic Killing vector fields, induced by flavor $U(1)$ rotations
on the soliton.

\subsection{counting generic BPS states}

This form of moduli dynamics with potential has well-known
supersymmetric extensions, provided that $K$ is a Killing vector field.
Such massive nonlinear sigma-model mechanics first appeared
with complex supersymmetry in a work by Freedman and Alvarez-Gaume
\cite{AlvarezGaume:1980dk}, while the form of relevance for us
was found more recently in the context of BPS dyons of the
Seiberg-Witten theory \cite{Gauntlett:1999vc,Gauntlett:2000ks,Weinberg:2006rq}.
In this subsection, let us outline this modified moduli dynamics and
solve for flavored BPS multi-kink states explicitly.
See appendix B for a short review.

Without the potential, the moduli dynamics
of the kinks would be the ordinary nonlinear sigma model where
the real fermions match 1-1 with real bosons. Therefore the
supercharge in question can be understood geometrically as
the spinorial Dirac operator on the moduli space,
\begin{equation}
{\cal Q}=i\Gamma^I\nabla_I \,,
\end{equation}
where $\nabla_I$'s are the covariant derivative with ordinary
spin connection and $\Gamma^I$'s the Dirac matrices.

The addition potential energy shifts this supercharge.
With general $L$-kink case,
the supercharge is shifted  as
\begin{eqnarray}\label{Q}
  {\cal Q} = \G^I \big( i\nabla_I + \sum_p m_E^{(p)} K_I^p \big) \,.
\end{eqnarray}
Taking square of this supercharge, one finds
\begin{equation}
\{{\cal Q},{\cal Q}\} ={\cal H}-{\cal Z} \,,
\end{equation}
where the central charge (to be distinguished from $Z$ of the field theory)
is defined via Lie-derivatives
\begin{equation}
{\cal Z}=-i\sum_p m_E^{(p)} {\cal L}_{K^p} \,,
\end{equation}
with respect to the Killing vectors, whose action is part of the
global $U(1)^N$ flavor rotations acting the kinks.

Since the BPS state must saturate the bound ${\cal H}-{\cal Z}=0$,
the search for BPS states in any given kink sector boils down to
finding zero modes of ${\cal Q}$ on the moduli space.
This task is in principle very complicated. However, one can reduce
counting problem to that of two-body problems, at least for the
index of such quantum mechanics.

With $m_E\neq 0$, the operator  ${\cal H}-{\cal Z}$ has a massgap
which separate the continuum from the ground state. Such operators
are called Fredholm operators, for which usual index theorem applies;
one simply choose to scale up the values of $a_p$'s, thus increasing
the mass gap indefinitely, while keeping the index unaffected.
This localizes the index computation to the fixed points of
the vector fields $K^a$'s. Once this happens, the counting problem becomes
that of harmonic oscillators and factorizes into minimal units with two bosonic
and two fermionic coordinates \cite{Stern:2000ie}. The latter is a
two-kink problem, so it suffices to count BPS bound states in a
two-kink problem in order to compute index for arbitrary multi-kink states.

For flavored 20-kink state problem, we have seen that the supercharge reduces to
\begin{eqnarray}
  {\cal Q} = \G^I \big( i\nabla_I + m_E K_I \big)\,,
\end{eqnarray}
when  $m_{20}=2m$ and $m_{10}=m+im_E$ with real $m$ and real $m_E$. The
Hamiltonian is nonnegative and has the general form
\begin{equation}
{\cal H}=\frac12(g_{rel})_{\mu\nu}\left(\pi^\mu \pi^\nu+m_E^2K^\mu K^\nu\right)+\cdots \,,
\end{equation}
where the ellipsis denotes terms involving fermions and $\pi^\mu$'s are
the canonical conjugate momenta of the moduli coordinates $y$'s.

With $\z_{rel}=e^{\rho+i\varphi}$, the metric for ${\cal M}_2$ is
\begin{equation}
g_\text{rel}= f(\rho)^2\left(d\rho^2+d\varphi^2\right)\,,
\end{equation}
where
\begin{equation}
f(\rho)^2\equiv \frac{2r}{m}e^{-4\rho}F(e^{4\rho})\,.
\end{equation}
In the relevant orthonormal frame,
\begin{eqnarray}
  e^{\hat \r} = f(\r)d\r\ , \ \ e^{\hat \varphi} = f(\r)d\varphi\ ,
  \ \ \omega^{\hat \varphi}_{\ \hat \r} = \frac{\partial_\r f(\r)}{f(\r)} d\varphi\,,
\end{eqnarray}
the supercharge reduces to
\begin{eqnarray}
  {\cal Q}  = \G^{\hat \r} \frac{1}{f(\r)} \bigg[
  \partial_\r + \frac12 \frac{\partial_\r f(\r)}{f(\r)}
  + i \G^{\hat \r \hat \varphi} \Big( q - m_E f(\r)^2 \Big)
  \bigg] \ , 
\end{eqnarray}
in the charge $q$ sector, that is, when $-i\partial/\partial\varphi\to q$.
A supersymmetric state in this sector has the central charge $qm_E$,
which must be saturated by the nonnegative Hamiltonian. Thus, a BPS bound
state is possible only if $qm_E\ge 0$.

Denoting two chiral components of $\Psi$ under
$i\G^{\hat \r \hat \varphi} $ by $u_{\pm}$, the
zero-mode solves
\begin{eqnarray}
  \partial_\r \big[\sqrt{f(\r)} u_\pm \big] \pm
  \Big( q - m_E f(\r)^2 \Big) \big[ \sqrt{f(\r)} u_\pm \big] = 0 \ ,
\end{eqnarray}
from which one can obtain one and only one normalizable solution
\begin{eqnarray}
  u_- = \frac{u_0}{\sqrt{f(\r)}} e^{iq\varphi}
  \text{exp}\bigg[
  \int^\r_{\r_0} d\r' \Big( q - m_E f(\r)^2  \Big) \bigg]\,,
\end{eqnarray}
whenever
\begin{eqnarray}
  0\leq q < q_\text{cr} = m_E f(\infty)^2 = 2r \frac{|m_\text{E}|}{m}
\end{eqnarray}
The upper bound comes from the asymptotic normalizability
while the lower bound is required by normalizability at origin ($\rho\to -\infty$),
\begin{eqnarray}
  u_- \simeq u_0 \left(\frac{8m}{r\pi}\right)^{\frac14} e^{(q-1/2)\r + iq\varphi}
  \exp{\bigg[-\frac{r|m_\text{E}|\pi}{16m} e^{2\r}\bigg]}\ .
\end{eqnarray}
Although $q=0$ wavefunction is mildly singular at origin, it is still
normalizable.\footnote{Note that the upper bound on the electric charge is precisely
the critical charge obtained from the classical construction of
flavored composite dyons in the section 3
\begin{eqnarray}
  q_{cr}=   Q_1 = -(Q_0 + Q_2) = r \big( \tan \th + \tan \tilde\th \big) \simeq
  2 r \frac{|m_\text{E}|}{m}\ .
\end{eqnarray}
}

In summary, we found exactly one flavored bound state of the 10-kink
and 21-kink for each integral relative charge $q$ from 0 up to
$q_{cr}=2r{|m_\text{E}|}/{m}$ and for arbitrary half-integral
(odd $q$) or integral (even $q$) $q_{cm}$. Each of such bound states
complete into a BPS multiplet, thanks to the Goldstino mode.
These flavored kinks become unstable against decay to a pair of
simple flavored kinks (10- and 21) when the mass parameters are
changed such that the critical relative charge $q_{cr}$ becomes
smaller or equal to $q$.

Index computation for more general flavored multi-kink states
follows immediately. As argued above, the problem factorizes into
several two-body problems. We consider general flavored
$L0$-kink, viewed as bound state of 10-kink, 21-kink, 32-kink, etc.
For the $p$-th pair, there is one ``relative'' flavor charge
$q^{(p)}$. When this charge obeys the conditions,
\begin{equation}\label{cc}
0\le |q^{(p)}| < q^{(p)}_{cr}(m_{i0})\;\;\;\hbox{and}\;\;\; 0< m^{(p)}_E q^{(p)} \,,
\end{equation}
the above two-body result tells us that the index is unit. The
total index for this $L$-body problem is a product of all such two-body
indices, so we learn finally that
\begin{equation}
\Omega=(-1)^f \,,
\end{equation}
where $f$ is the $R$-charge of the soliton,
provided that (\ref{cc}) is satisfied for all $p=1,2,\dots,L-1$.
Otherwise
\begin{equation}
\Omega=0 \,,
\end{equation}
which we will take as an evidence that the corresponding BPS does not
exist.

\subsection{wall-crossing }

After lengthy computations, we finally arrive at wall-crossing issues
at large mass limit of this massive $D=2$ QED.
Since $q^{(p)}_{cr}\sim rm_E^{(p)}/m_{L0}$, there is a wall of marginal
stability for these flavored kink at $rm_E^{(p)}/m_{L0}\sim q^{(p)}$,
details of which would follow once we compute the metric and the
potential on the moduli space. This is a tedious but straightforward
exercise. For us, it suffices to know that these walls of marginal stability
are determined by $r$ and $q$'s, and they extend to the asymptotic
region of large $r$. Across any such a wall, the flavored multi-kink
states break into a pair of smaller flavored multi-kink states, such as
$L0$-kink interpolating between 0 vacuum and $L$ vacuum breaking
up into a flavored $K0$-kink and a flavored $LK$-kink.
The latter two
objects exist on both side of this particular wall, so the jump in the
spectrum is only for the bound state, and we have the simple jumping
formula
\begin{equation}
|\Delta\Omega |=1 \,.
\end{equation}
As we saw in section 2, the marginal stability wall is, as always, defined by
the phase alignment of the two central charges of the flavored $K0$-kink
and the flavored $LK$-kink.

In fact, this simple wall-crossing formula is a special case of general
wall-crossing where we are considering bound states of two BPS particles
with unit degeneracy. For this, let us review a result from \cite{Cecotti:1992qh}.
They defined a twisted partition function of $D=2$ field theories as
\begin{equation}
{\cal F}(\beta;m^i)=\lim_{l\to \infty}\frac{i\beta}{l}\tr(-1)^R R e^{-\beta H} \,,
\end{equation}
where $l$ is the regulated size of the spatial line. Alternatively
this may be thought of as expectation value of $R$ when the theory
is defined on $S^1\times {\mathbb R}^1$ with Euclidean signature
and periodic boundary condition on $S^1$. A single-particle
BPS state, $Z$, contributes
\begin{equation}
{\cal F}_{Z}=i\beta (-1)^f \int\frac{dp}{\pi}e^{-\beta\sqrt{p^2+|Z|^2}}
=\frac{i(-1)^f}{\pi}\int{d\mu}\;\beta |Z|\cosh\mu \;e^{-\beta|Z|\cosh\mu} \,,
\end{equation}
with the rapidity $\mu=\sinh^{-1}(p/|Z|)$.
Note that, as we vary the parameters of the theory, wall-crossing will
occur somewhere and this contribution from single particle BPS states
will have to be disappear in a discontinuous manner.

On the other hand, $\hat \Omega$ also receives contributions from
many particle sectors. In particular, with the decomposition of
the central charge as, $Z=Z_1+Z_2$, the two-particle contribution
is of some interests. Following
Cecotti et.al., we also finds that, when the pair of BPS states
$Z_{1,2}$ backscatter,\footnote{Even in $D=2$ what one means by forward-scattering
and backward-scattering can be somewhat ambiguous when particles can change species. However, we are
mostly interested in situations when two particles in question are clearly
distinct, with different masses for example, so that the particles are
unambiguously labeled. In this context, backscattering
means the sign flip of the relative rapidity before and after.}
there is a contribution from the two-particle sector of the type
\begin{eqnarray}
&&{\cal F}_{Z_1+Z_2}=\nonumber \\ \nonumber \\
&&d_{{1}}d_{{2}}\frac{i(-1)^{f_1+f_2}}{4\pi^2}\int\int{d\mu_1} d\mu_2\;
\beta\left( |Z_1|\cosh\mu_1+ |Z_2|\cosh\mu_2\right)e^{-\beta(|Z_1|\cosh\mu_1+|Z_1|\cosh\mu_1)}\nonumber\\
&&\hskip 4cm \times \frac{\partial}{\partial\mu_1}\log\left(\sinh(\mu_2-\mu_1+i\epsilon)/\sinh(\mu_1-\mu_2+i\epsilon)\right) \,,
\end{eqnarray}
where $2\epsilon={\rm Im}\log(Z_2/Z_1)$ and $d_{{1,2}}$ are the number of
such BPS supermultiplets of central charge $Z_{1,2}$.

Recall that the wall of marginal stability would be at $\epsilon=0$ where
the two central charges line up in the complex plane. Because of the logarithm, the two-particle
expression ${\cal F}_{Z_1+Z_2}$ also has a discontinuous imaginary part,
and in fact
\begin{equation}
\lim_{\epsilon\to 0^\pm} {\cal F}_{Z_1+Z_2}=\pm d_{{1}}d_{{2}}\frac{{\cal F}_{Z}}{2} \,,
\end{equation}
so that
\begin{equation}
\lim_{\epsilon\to 0^+} {\cal F}_{Z_1+Z_2} -\lim_{\epsilon\to 0^-} {\cal F}_{Z_1+Z_2}=d_{{1}}d_{{2}}{\cal F}_{Z} \,.
\end{equation}
Although individual contributions are discontinuous,
the twisted partition function  $\hat \Omega$ itself can
be continuous provided that $Z$ state exists as a one-particle
BPS state only on the $\epsilon <0$ side. The continuity
of the twisted partition function seems reasonable, and
this would then imply a rather general wall-crossing behavior.
Assuming such a continuity of ${\cal F}$, and since $\Omega(Z_{1,2})
=(-1)^{f_{1,2}}d_{{1,2}}$, we the find the general
wall-crossing formula across $Z\rightarrow Z_1+Z_2$ walls of marginal stability,
\begin{equation}\label{2dwc}
\Delta\Omega(Z)=\pm \Omega(Z_1)\Omega(Z_2) \,.
\end{equation}
For flavored domain walls in the massive ${\mathbf CP}^N$ theory,
we found $|\Delta\Omega(Z)|=1$, which is easily explained by this
wall-crossing formula,  since elementary excitations and simple
kinks all have unit index, $|\Omega|=1$. Building
more complicated flavored kinks out of them  can only generate
flavored kinks with $|\Omega|=1$ because the wall-crossing formula
(\ref{2dwc}) is so simple.

Wall-crossing in $D=2$ was originally studied by Cecotti and Vafa
for purely topological kinks \cite{Cecotti:1992rm}. For this case,
the central charges simplifies as differences of ``canonical
coordinates" which in our case are simply the masses $m^i_D\simeq
\tau m^i$, and ${\cal F}$ can be explicitly solved using
the $tt^*$ equations \cite{Cecotti:1991me}. Introduction of
flavor charges to the kink should modify the latter approach
somewhat, if not drastically, which will appear elsewhere.

\section{$D=4$ $\CN=2$ $SU(N+1)$ with flavors}

This two-dimensional QED shows certain features reminiscent of
the Seiberg-Witten theory of four dimensions. This was first noted
by Hanany and Hori \cite{Hanany:1997vm} who found that the
renormalization of the FI parameters $\tau=-ir+\theta/2\pi$
and the asymptotic form of the four-dimensional $\tau_{SW}$
have a close resemblance.
This was taken up later more seriously by Dorey \cite{Dorey:1998yh}
who argued that the spectrum of this theory is related to that of
$SU(N+1)$ Seiberg-Witten theory with $N+1$ flavors of
masses $m^i$. The correspondence was supposed to be precise at the root
of the baryonic branch where the vacuum expectation values of the
Seiberg-Witten scalars match with the quark masses. This conjecture
was further extended by Dorey, Hollowood, and Tong \cite{Dorey:1999zk}.

The most compelling reason for this conjecture comes from the
exact central charge (\ref{central1}) of the BPS states, obtained from
effective superpotential $\CW(\S)$ after integrating over all chiral
multiplets of (\ref{GLSMlag2}) in the parameter region $e\ll \L_\s$.
In \cite{Hanany:1997vm}, it has been pointed out that the periods
$m_D^i-m_D^j$ (\ref{central3}) are in perfect matching with those of
the Seiberg-Witten curve at baryonic root of the corresponding
$D=4$ $\CN=2$ $SU(N+1)$ gauge theory with massive $N+1$ quarks.

This latter observation, strictly speaking, tells
us only that the set of central charges in the two theories
may coincides, not necessarily the actual particle content.
Nor does not say much about degeneracies of general BPS states on
the two sides. Yet, one may go a bit further and hope that at least
hypermultiplets of
Seiberg-Witten theory may match against $D=2$ spectra, since these
can be potentially massless somewhere in the moduli space (or parameter
space for $D=2$) and can be associated with singular structure of the
latter. This is precisely the conjecture of Dorey and his collaborators.

Now that we found a very rich spectrum of flavored kinks,
counted their degeneracy, and found the wall-crossing behavior,
let us come back to this conjecture and see how it lives up to its
promise.
In generic Seiberg-Witten theory of rank large than one, typical
BPS dyons are not in the hypermultiplet. Rather they come with large
angular momentum which is already evident in the classical soliton
solutions. As we will see below, under the proposed correspondence
between $D=2$ QED and the Seiberg-Witten theory, a typical flavored kink
we found would be mapped to such dyons with high angular momenta.
Let us explore to what extent and in what sense there might be
an``equivalence" of BPS spectra of the two theories.

Recall the central charge of Seiberg-Witten theory,
\begin{equation}
Z_{SW}=\vec a_D\cdot \vec{Q}_m+\vec a \cdot \vec{Q}_e +\sum_f m^fS_f \,.
\end{equation}
In the asymptotic region, we have $\vec a_D=
\tau_{4D} \vec a$. For $SU(N+1)$ theory with $N+1$ fundamental
hypermultiplets, we have a special point where $a^i=m^{f=i}$, where
the central charge simplifies to
\begin{equation}\label{C}
Z_{SW}=\tau_{4D} \vec m\cdot \vec{Q}_m+\vec m \cdot \vec{Q}_e^{adj}+ \sum (m^i-m^j)\tilde Q_{ij} \,.
\end{equation}
$ Q_e^{adj}$ denotes electric charges in the adjoint
root lattice and the combined contribution from the matter multiplet
\begin{equation}
\tilde Q=S+Q_e^{matter}
\end{equation}
effectively lives in a $SU(N+1)$ root lattice, which explains
why we wrote the last term in Eq.~(\ref{C}) as mass differences.
For ``unit" magnetic charges,
we have the following mapping from $D=4$
theories,
\begin{eqnarray}\label{2d4d}
Q_m &\rightarrow &T \,,\nonumber \\
Q_e^{adj}+\tilde Q &\rightarrow& Q\,,\nonumber\\
\tau_{4D} &\rightarrow & \tau=\frac{\theta}{2\pi}-ir \,,\nonumber\\
\big(\vec a, \vec a_D\big) &\rightarrow& \big( \vec m , \vec m_D \big)\,,
\end{eqnarray}
to $D=2$.
Note that $Q$'s we found are always in the root lattice
which is achieved on the left hand side by mixing of $SU(N+1)$
color weights and $SU(N+1)$ favor weights at this special
point in the Seiberg-Witten moduli space
This map forms the basis
of the conjectured equivalence of BPS spectra on the two sides.
Writing the root system of $SU(N+1)$ as collection of $e_i-e_j$
with $0\le j<i\le N$, and mapping the $D=2$ central charge to
this, we see that the $ki$-kink corresponds to a magnetic root
of $e_k-e_i$ whereas $jl$ flavor charge maps to either
a $(e_l-e_j)$-vector meson, or an $e_j$ colored quark of $l$-th
flavor (or vice versa).

Finally, the relevant index for $D=4$ $\CN=2$ theory
is the second helicity trace.
\begin{equation}
\Omega_{SW}=-2\,\tr (-1)^F J_3^2 \,.
\end{equation}
which counts various BPS multiplets with some weights.
Actual values are
\begin{equation}
\Omega_{SW}([s]_{spin}\otimes[{\rm half\;Hypermultiplet}])=(-1)^{2s}(2s+1)\,,
\end{equation}
where the first factor denotes the angular momentum
multiplet under the $SO(3)$ little group, denoted by its spin.
For example, a charged vector gives $-2$.

\subsection{BPS dyons in pure $SU(N+1)$ and wall-crossing }

What are known in literature about such a large-rank Seiberg-Witten
theory come from weak coupling analysis, that is, in the limit of
large vacuum expectation values.\cite{See Ref.~\cite{Weinberg:2006q}
for a comprehensive review.} In this regime, the low energy
dynamics of monopoles are easily set up and reliable for general
$\CN=2$ theories. In particular, dyons in pure $SU(N+1)$ theory whose magnetic
charge is a (dual) root, as opposed to arbitrary linear combinations
thereof, are completely classified and counted by Stern and Yi \cite{Stern:2000ie}.
 Let us summarize their result first.

As in $D=2$, an ordering is possible when the
adjoint vacuum expectation values $a^i=m^i$ almost line up in the
complex plane. By overall $U(1)$ rotation, we can take them to be
almost real,  such that
\begin{equation}
{\rm Re} \,m^0 <{\rm Re}\,m^1 <\cdots {\rm Re}\,m^{N} \,,
\end{equation}
as we did in the previous sections for $D=2$ theory.
Without loss of generality, take dyons of magnetic
charge $e_L-e_0$. With the above ordering of vev's, electric charges
of dyons are restricted as
\begin{equation}
-\left(\frac{k+\sum n^{(p)}}{2}\right)e_0+ n^{(1)}e_1+n^{(2)}e_2+
\cdots +n^{(L-1)}e_{L-1} +\left(\frac{k-\sum n^{(p)}}{2}\right)e_L \,,
\end{equation}
with integers $k$ and $n^{(p)}$'s correlated such that the
coefficients of $e_{L,0}$ are also integral. For a BPS dyon
of such a charge to exist, the charges must obey the inequalities
\begin{equation}
n^{(1)}\times {\rm Im}\, m^1 >0 ,\quad n^{(2)}\times {\rm Im}\, m^2 >0 ,
\quad \dots,\quad n^{(L-1)}\times {\rm Im}\, m^{L-1} >0 \,,
\end{equation}
and also that the individual electric charge does not exceed the
critical value, which goes as
\begin{equation}
|n^{(p)}| < \frac{8\pi^2}{e^3}\sum_q \mu^{-1}_{pq}{\rm Im}\,m^q \,,
\end{equation}
where the matrix $\mu$ is a reduced mass matrix defined
in terms of ${\rm Re}\, m^q$'s. See Ref. \cite{Stern:2000ie,Weinberg:2006rq}

When these conditions are satisfied, the degeneracy is known \cite{Stern:2000ie}.
Furthermore, the angular momentum content is also not difficult to find, and the end
result is that the dyon is in the following multiplet,
\begin{equation}
\left(\otimes_p \left[\frac{|n^{(p)}|-1}{2}\right]\right)\otimes  [{\rm half\, Hypermultiplet}] \,.
\end{equation}
Note that the dyon appears not as a single supermultiplet but rather
as a sum of many supermultiplets with spins up to $(\sum |n^{(p)}|-L+1)/2$.
The index $\Omega_2$ of such a dyon is
\begin{equation}
\Omega_{SW}= (-1)^{\sum n^{(p)}-L+1}\prod_p |n_{(p)}| \,.
\end{equation}
In fact, the computation of BPS bound states for kinks of previous
section is modeled after the computation here. This result was
later reproduced by Denef from more stringy viewpoint \cite{Denef:2002ru}.

Recently a startling proposal by Kontsevich and Soibelman (KS) \cite{Kontsevich:2008ab}
was given for all wall-crossing behavior of $D=4$ $\CN=2$ theories, which
seems to fit all known examples of wall-crossings of these theories.
For our purpose, we will not really need the full power of KS proposal
but a corollary for the so-called semi-primitive cases. One considers BPS
bound states of the form $\gamma(s)=\gamma_1+s\gamma_2$, where $\gamma$'s
denote electromagnetic charges of the states and we assume that $\gamma_{1,2}$
are primitive, namely they are not integer multiple of other charge vector.
Denoting $\Omega_{t,s}\equiv \Omega_{SW}(t\gamma_1+s\gamma_2)$, we have
the wall-crossing formula for $\Omega_{1,s}$ as a consequence of KS
formula;
\begin{equation}
\Omega_{1,0}+\sum_{s\ge 1} \Delta \Omega_{1,s}y^s= \Omega_{1,0}\prod_{s'\ge 1}
\left(1-(-1)^{s'\langle\gamma_1,\gamma_2\rangle}y^{s'}\right)^{\pm s'\langle \gamma_1,\gamma_2\rangle\Omega_{0,s'}} \,.
\end{equation}
The Schwinger product of the charges $\langle \gamma_1,\gamma_2\rangle$
enters the exponents everywhere. When only $\Omega_{t,0}$ and $\Omega_{0,s}$
are nonzero on one side of the wall, this would determine $\Omega_{1,s}=
\Delta \Omega_{1,s}$ completely on the other side of the wall.

This was first suggested by Denef and Moore \cite{Denef:2007vg}
as a phenomenological formula. It can also be derived from the
KS formula, which shows how to fix the sign in the last exponent
in terms of the sign of the relative phase of the two central
charges $Z_1$ and $Z_2$ on the side of the wall.
We left the sign ambiguous since we will presently fit this formula to
the known spectrum where the correct sign appears quite obviously.

A further simplification results if we take $\Omega_{0,s}=0$
for all but $s=1$. As far as we know, in all  $D=4$
$\CN=2$ field theories, no non-primitive charge state has ever
been found as one particle states.\footnote{This is one notable
difference from the supergravity countings, despite many other
similarities. We do not know of an explicit proof of this statement,
although there were examples where this absence was shown in some cases.}
Then we have,
\begin{equation}
\Omega_{1,0}+\sum_{s\ge 1} \Omega_{1,s}y^s= \Omega_{1,0}
\left(1-(-1)^{\langle\gamma_1,\gamma_2\rangle}y\right)^{\pm \langle
\gamma_1,\gamma_2\rangle\Omega_{1,0}} \,.
\end{equation}
Let us see how this fits with the known spectrum of dyons
we discussed above. Take for example the simplest $L=2$.
We will write the charge vectors as $\gamma_1= (e_2-e_0; e_1-e_0)$
and $\gamma_2= (0; e_1-e_0)$ so that
\begin{equation}
\gamma(s)=(e_2-e_0; (s+1)e_1-(s+1)e_0) \,.
\end{equation}
In terms of dyons whose degeneracy we saw earlier,
this corresponds to $L=2$, $n^{(1)}=k=s+1$. One may be tempted to
take $\gamma_1= (e_2-e_0; 0)$ but this state is absent in this
corner of moduli space and cannot be used as $\gamma_1$.

{}From
the knowledge of $\Omega_{1,0}=1$ and $\Omega_{0,1}=-2$
(because it is a vector multiplet), we find
\begin{equation}
\sum_{n\ge 0}y^s\Omega((e_2-e_0; (s+1)e_1-(s+1)e_0)) =
\left(1+y\right)^{\pm 2} \,.
\end{equation}
With the negative sign in the exponent (which is
something that can be checked independently), we find
\begin{equation}
\Omega_{SW}((e_2-e_0; n^{(1)}e_1-n^{(1)} e_0)) = (-1)^{n^{(1)}-1}n^{(1)} \,,
\end{equation}
after putting $n^{(1)}=s+1$ in the expression.
It is clear that this procedure can be repeated for more
complicated dyons with $L>3$ by taking $\gamma_2=(e_p-e_0)$ for
all $p=1,\dots,L-1$, which  results in
\begin{equation}
\Omega_{SW}((e_L-e_0; \sum_{p=1}^{L-1} n^{(p)}e_p
-\sum_{p=1}^{L-1} n^{(p)} e_0)) = (-1)^{\sum (n^{(p)}-1)}\prod_p n^{(p)} \,,
\end{equation}
in precise accordance with the general index formulae
computed in the low energy dynamics approach. Now that
we have some confidence in how wall-crossing formula
reproduce known spectra, let us move on to the flavored cases.

\subsection{flavored dyons from wall-crossing formula }

The actual dyons whose spectra was proposed to be equivalent
to that of $D=2$ theory are those that appear in $SU(N+1)$
Seiberg-Witten theory with $N+1$ fundamental hypermultiplets
with masses $m_i$'s. Furthermore, the comparison can be made
only at the root of the baryonic branch. Recall that well inside
the baryonic branch, where electric charges are screened,
the vector mesons and massive hypermultiplets together form
a long multiplet. Let us denote them as
\begin{equation}
W_{ij},\;\;q_i^{(j)},\;\;\tilde q_j^{(i)},
\end{equation}
where $q$, $\tilde q $ are the two chiral multiplets
of the hypermultiplets and are, respectively, in
the representations $(N+1, \overline{N+1})$ and $(\overline{N+1}, {N+1})$
under $SU(N+1)_{gauge}\times SU(N+1)_{flavor}$.
Given the map (\ref{2d4d}),
the correspondence between the flavored kinks and $D=4$ dyons
are easy to see.

Let us first consider the simplest nontrivial case with $L=2$.
The kinks of topological and flavor charge\footnote{Although
general flavored kink in this simple example would be more like
$$(T,Q)=(e_2-e_0;k'(e_2-e_0)+n(e_1-e_0))$$
for any integer $k'$, we set $k'=0$ because it  affects
neither the marginal stability nor degeneracy, at least in the
leading order in $1/r$. The same goes for $L0$-kink cases we
later consider.}
\begin{equation}
(T,Q)=(e_2-e_0;n(e_1-e_0))\,,
\end{equation}
can be mapped to a monopole of charge $(e_2-e_0)$, which we denote by $M_{20}$,
bound with $n$ electrically charged particles which can be
either $W_{10}$ or $\tilde q_0^{(1)}$. The other quark,
$\tilde q_1^{(0)}$ cannot bind to this monopole since
it does not have the right dynamical charge. Thus we find
the following map,
\begin{equation}
(T,Q)=(e_2-e_0;n(e_1-e_0)) \leftarrow
M_{20}+nW_{10} \;\; or\;\; M_{20}+(n-1)W_{10}+  \tilde q_0^{(1)} \,.
\end{equation}
The quark cannot bind more than once due to the Pauli exclusion
principle, although this can also be deduced from the wall-crossing
formula. See below.

In figuring out degeneracies of these dyons, one crucial
information missing is with what minimal electric charge the dyon
actually exist as a hypermultiplet. In this asymptotic corner
and in the pure $SU(N+1)$ case, we saw that
$M_{20}+W_{10}$ is the first such hypermultiplet. With flavors present,
this need not be true anymore. In fact the original conjecture on
equivalence of $D=2$ and $D=4$ spectra relied heavily on the fact
that the two theories share the same spectral curve, suggesting
that at least hypermultiplet content of $D=4$ theory should be
faithfully reflected in $D=2$ theories. This leads us to guess
that the first hypermultiplet is the purely magnetic bound state,
$M_{20}$, namely a magnetic monopole of charge $e_2-e_0$.
Our objective here is to reproduce the rest of BPS spectra
from this single assumption.

 We may naively repeat the analysis of the pure case. From the
 wall-crossing formula, we deduce that
\begin{eqnarray}
\sum y^s\Omega_{SW}(M_{20}+s \tilde q_0^{(1)}) = 1+y \,,
\end{eqnarray}
which, as promised, shows that quarks can bind to a monopole
at most once. Using the wall-crossing formula one more time, we find
\begin{eqnarray}
&&\Omega_{SW}(M_{20}+nW_{10} ) = (-1)^{n}(n+1) \,,\nonumber\\
&&\Omega_{SW}(M_{20}+(n-1)W_{10} +  \tilde q_0^{(1)}) = (-1)^{n-1}n \,.\nonumber
\end{eqnarray}
Note that individual spectra of these dyons are rather nontrivial
and come with high angular momentum content. However,
tt is intriguing that the sum of these two indices is rather simple
\begin{equation}
\Omega_{SW}(M_{20}+nW_{10} ) +\Omega_{SW}(M_{20}+(n-1)W_{10} +  \tilde q_0^{(1)}) = (-1)^{n} \,,
\end{equation}
and actually coincides with the $D=2$ counting of flavored kinks,
up to a sign.

More generally, for dyons with magnetic charge $e_L-e_0$, the
relevant indices are
\begin{eqnarray}
\Omega_{SW}(M_{L0}+ \sum_{p=1}^{L-1} l^{(p)} W_{p0} +  \sum_{p'}\tilde q_0^{(p')})
= (-1)^{\sum l^{(p)}}\prod (l^{(p)}+1) \,,
\end{eqnarray}
where $\{p'\}$ is a subset of $\{1,2,\dots,L-1\}$.
The map to $D=2$ flavored kink follows the same rule as before;
These dyons are mapped to flavored $L0$-kinks with $p0$-flavor charges
$q^{(p)}$ being equal to either $n^{(p)}=l^{(p)}$ (when $p\neq p'$) or $n^{(p')}=l^{(p')}+1$. Summing
over the indices for fixed $q^{(p)}=n^{(p)}$'s, we find
\begin{eqnarray}
\sum_{\{p'\}}
\left(\prod_{p=1,p\neq p'}^{L-1} (-1)^{ n^{(p)}}
(n^{(p)}+1)\prod_{p'} (-1)^{n^{(p')}-1} (n^{(p')})\right) \,,
\end{eqnarray}
which is the same as
\begin{eqnarray}
(-1)^{\sum n^{(p)}}\prod_{p=1}^{L-1}
((n^{(p)}+1)-n^{(p)})= (-1)^{\sum n^{(p)}} \,.
\end{eqnarray}
We thus find that under the proposed map (\ref{2d4d}), $D=2$ indices
equal precisely to the sum of $D=4$ indices of all corresponding
dyons, possibly up to a sign.

Note that this cancellation among $D=4$ indices, and the resulting match
against $D=2$ index, is possible only upon very fine-tuned relationships
among these dyons with different quark contents.

\section{Conclusion}

In this paper, we reviewed $D=2$ $\CN=(2,2)$ QED with twisted masses, with
emphasis on BPS spectra in the large mass limit. With $N+1$ chiral
matter fields, one finds BPS kink solutions endowed with $U(1)^N$
flavor charges, whose stability criteria mimics those of $D=4$ $\CN=2$
dyons. In the classical limit, this also coincides with that of open
string web, or equivalently 1/4 BPS dyons of $\CN=4$ Yang-Mills theory,
giving us a pictorial way to determine the marginal stability
walls. We quantized these solitons to obtain degeneracies, which
turned out to be unit for all such solitons. This result is
consistent with general wall-crossing behavior expected in
$D=2$ $\CN=(2,2)$ theories, namely,
$$
\Delta\Omega(Z_1+Z_2)=\pm \Omega(Z_1)\Omega(Z_2) \,.
$$
Wall-crossing of $D=2$ topological kinks has been studied
in depth where $tt^*$ equation makes a prominent appearance.
It would be very interesting to explore further
how this could be refined  to situations
with conserved charges (such as flavor charges) other than
topological charges.

We also compared the spectrum to the conjectured $D=4$ counterpart,
i.e., that of the $SU(N+1)$ Seiberg-Witten theories with $N+1$
massive fundamental hypermultiplets, at the root of the baryonic
branch. Due to the special nature of this point in the moduli space,
where the gauge symmetry and the flavor symmetry are locked,
one type of flavored kink is mapped to several different kind of
dyons with different quark contents. The degeneracies of the latter,
as counted by the second helicity trace, can be complicated
and large unlike those of the kinks. However, this difference is remedied
miraculously once we sum over the indices of all the corresponding
dyons with different quark content, which gives at the end,
\begin{equation}\label{=}
|\Omega|=1=|\sum_{\rm dyons} \Omega_{SW}| \,,
\end{equation}
for each flavored kink that exists on the left hand side and
for all the corresponding dyons on the right hand side.

One cannot really say that
spectra of the two theories are equivalent, since
various dyons that are mapped to one type of flavored kink
will generally carry
mutually different electric and flavor charges.
Note also that in this map only a subset of $D=4$ BPS dyons
participate. A topological charge of a kink is always
mapped to a dual root of the gauge group; since general
dyons may carry more general (magnetic) weight that lie
in the dual root lattice, there must be dyons that do not fit
in this correspondence. Given such obvious differences,
the agreement (\ref{=}) is all the more remarkable.

The question of whether and how wall-crossing behaviors and indices
of $D=2$ theories and those of $D=4$ theories might be related
 deserves further study. $D=4$ wall-crossing received much attention lately,
 as we noted already, and some of mathematical tools there
 have uncanny resemblance to those of $tt^*$ equations.
Whether such a mathematical resemblance has anything to do
with the present example is unclear, but it still
begs for a clarification. In particular,
the partial agreement (\ref{=}) of $D=2$ and $D=4$ indices,
despite vastly different BPS spectra with their different-looking
individual indices, needs to be understood better.
In a recent study \cite{Gaiotto:2009fs}, Gaiotto
pointed out a relationship between surface operators
in $D=4$ $\CN=2$ gauge theories and $D=2$ sigma model whose
UV theory is $\CN=(2,2)$ QED with massive chiral matters. It
would be  interesting to see what are the implications
in the present context.

\vskip 2cm

\centerline{\large \bf Acknowledgement}
\vskip 5mm

We thank Kentaro Hori, Yoon Pyo Hong, Seok Kim, Ki-Myeong Lee,
Sangmin Lee, and Jaemo Park for valuable discussions.
P.Y. thanks Yukawa Institute of Theoretical Physics and organizers
of the workshop,``Branes, Strings, and Black Holes'' for hospitality.
P.Y. is also grateful to the Center for Theoretical Physics, Seoul
National University, where part of this manuscript was written. P.Y. was
supported in part by the National Research Foundation of Korea(NRF)
grant funded by the Korea government(MEST) (No. 2005-0049409).

\newpage

\centerline{\Large \bf Appendix}

\appendix

\section{Miscellany}

\paragraph{notations and conventions}

One convenient way to describe two-dimensional supersymmetric
theories is to use the four-dimensional superspace formalism
of Wess and Bagger followed by a suitable dimensional reduction:
let us compactify the four-dimensional theories along $x^1, x^2$
directions so that chiral and anti-chiral spinors $\psi_\a, \bar \psi_\da$
reduce to two-dimensional complex spinors
\begin{eqnarray}
  \big( \psi_{1} , \psi_{2} \big) \equiv \big( \psi_{+},
  \psi_{-} \big),
  \qquad
  \big( \bar \psi_{\dot 1} , \bar \psi_{\dot 2} \big)
  \equiv \big( \bar \psi_{-}, \bar \psi_{+} \big)\ .
\end{eqnarray}
Here $\pm$ denote the charges under $U(1)_\text{A}$
R-symmetry, arising from the spatial rotation in the
compactified dimensions.

In addition to usual superfields with four supercharges
such as vector and chiral superfields, it is well-known
that two-dimensional theories allow a so-called twisted
chiral superfield. The twisted chiral superfield $\hat \Phi$ is
defined as
\begin{eqnarray}
  \bar D_+ \hat \Phi = D_+ \hat \Phi = 0 \ .
\end{eqnarray}
Defining twisted fermionic coordinates
$\hat \th_\a= ( \th_+ , - \bar \th_+ )$, the twisted chiral
superfield has the following component field expansion
\begin{eqnarray}
  \hat \Phi = \hat \phi + \sqrt2 \hat \th \hat \psi
  + \hat \th \hat \th \hat F\ .
\end{eqnarray}
As a comment, the chiral/twisted chiral-multiplets are indeed
in a mirror pair.

One peculiar example of such twisted chiral superfields
is of the form
\begin{eqnarray}
  \S = D_+ \bar D_+ V \ ,
\end{eqnarray}
where $V$ denote the vector multiplet.
The component field expansions of the above superfield $\S$
read
\begin{eqnarray}
  \S &=& \big(A_1 - i A_2\big) + 2i
  \bar \th_+ \l_+ + 2i \th_+ \bar \l_+ + 2\th_+ \bar \th_+
  \big(  D + i F_{03} \big) + \cdots \nn \\
  &=& \hat \phi + \sqrt 2 \hat \th \hat \psi + \hat \th \hat \th
  \hat F
\end{eqnarray}
with
\begin{eqnarray}
  \hat \phi = A_1 - i A_2, \qquad \hat \psi_\a = - \sqrt2 i
  \big( \l_+, \bar \l_+ \big), \qquad \hat F =  D + i F_{03}
  \nonumber \ .
\end{eqnarray}
Using $\S$, the Fayet-Iliopoulos term and topological $\th$-term
can be combined as
\begin{eqnarray}
  \CL_\text{FI} + \CL_\th=
  - \text{Im} \Big[ \t \int d^2 \hat\th \  \S \Big]
  = r D - \frac{\th}{2\pi}F_{03}\ ,
\end{eqnarray}
where $\t= - i r + \frac{\th}{2\pi}$.


\paragraph{covariant derivative}

Using the inhomogeneous parameterization $z^m$ of $\mathbb{CP}^{N}$,
the GLSM scalar fields can be expressed up to overall $U(1)$ phase
as
\begin{eqnarray}
  \phi^0 = \sqrt{\frac{r}{1+ \bar z_m z^m}}\ , \qquad
  \phi^n = \sqrt{\frac{r}{1+\bar z_m z^m}} z^n\ .
\end{eqnarray}
The $U(1)$ gauge field $A_\mu$  (\ref{gauge}) now in turn becomes
\begin{eqnarray}
  A_\mu = \frac{\bar z_m \partial_\mu z^m - \partial_\mu \bar z_m z^m}
  {2i\big(1+\bar z_m z^m \big)}.
\end{eqnarray}
The various covariant derivatives are then given by
\begin{eqnarray}
  D_\mu \phi^0 &=& - \sqrt{\frac{r}{1+ \bar z_m z^m}} \
  \frac{\bar z^m \partial_\mu z_m}{ 1+ \bar z^m z_m }\ ,
  \nonumber \\
  D_\mu \phi^n &=& + \sqrt{\frac{r}{1+ \bar z_m z^m}} \ \Big[
  \partial_\mu z^n - \frac{z^n \big(\bar z^m \partial_\mu z_m\big)}
  { 1+ \bar z^m z_m}\Big]\ .
\end{eqnarray}
Inserting the above results back into the BPS equation (\ref{BPSeqn}),
one can obtain (\ref{timeBPSeqn}).

\paragraph{energy for composite kinks}

For the composite kink solution, it needs
much elaboration to massage the energy functional to sum of
complete squares and boundary terms.
Since two mass parameters $m_{10}$ and $m_{20}$ are now parallel,
let us set them to be purely real without loss of generality.
From the general expression of energy functional (\ref{energy}),
one can obtain
\begin{eqnarray}
  \CE &=& \hspace*{-0.3cm}  \int d{\bf x}^3 \
  \frac{r}{\big( 1+ |z^1|^2 + |z^2|^2\big)^2} \left[
  \frac{(1+|z^1|^2 + |z^2|^2)
  \big|\bar z_2 \partial_3 z_1 - \bar z_1 \partial_3 z_2 \big|^2}{|z^1|^2+|z^2|^2}
  \right.
  \nonumber \\ && \hspace*{3.5cm}
  + \frac{\big| \bar z_1 \partial_3 z^1 + \bar z_2 \partial_3 z^2 \big|^2}{|z^1|^2+|z^2|^2}
  + m_{10}^2 |z^1|^2 + m_{20}^2 |z^2|^2 + m_{12}^2 |z^1|^2|z^2|^2 \Bigg]
  \nonumber \\
  &=& \hspace*{-0.3cm}  \int d{\bf x}^3 \
  \frac{r}{\big( 1+ |z^1|^2 + |z^2|^2\big)^2} \Bigg[
  \frac{\big| \bar z_1 (\partial_3 z^1 -m_{10} z^1 )
  + \bar z_2 (\partial_3 z^2 - m_{20}z^2) \big|^2}{|z^1|^2+|z^2|^2}
  \nonumber \\ &&  \hspace*{3.5cm}
  + \frac{\big|\bar z_2 (\partial_3 z^1 - m_{10}z^1)
  - \bar z_1 (\partial_3 z^2 - m_{20} z^2 ) \big|^2}{|z^1|^2+|z^2|^2}
  \nonumber \\ && \hspace*{3.5cm}
  + \big( m_{10} + m_{12} |z^2|^2 \big) \partial_3 |z^1|^2
  + \big( m_{20} - m_{12} |z^1|^2 \big) \partial_3 |z^2|^2 \Bigg]
  \nonumber \\ &\geq& \left.  \frac{r}{1+|z^1|^2+|z^2|^2}
  \Big( m_0  + m_1 |z^1|^2 + m_2 |z^2|^2 \Big)
  \right|^{{\bf x}^3=+\infty}_{{\bf x}^3=-\infty} = r m_{20}\ .
\end{eqnarray}
It implies that the composite kink saturating the bound
has the same mass as the simple $(20)$-kink solution.

\section{Low energy dynamics of kinks}

\subsection{fermion zero mode counting with aligned masses}


We begin by clarifying the number of fermionic zero modes in the simple kink background.
Under the $(20)$-kink background,
one can naturally define inner products of $\chi^{1,2}$ as
\begin{eqnarray}
  \langle \tilde \chi^1 | \chi^1 \rangle &=& \int d{\bf x}^3 \
  \frac{1}{1+e^{2|m_{20}|{\bf x}^3}} \tilde \chi^{1\dagger} \chi^{1}\ ,
  \nonumber \\
  \langle \tilde \chi^{2} | \chi^{2} \rangle &=& \int d{\bf x}^3 \
  \frac{1}{\big(1+e^{2|m_{20}|{\bf x}^3}\big)^2} \tilde \chi^{2\dagger} \chi^{2}\ ,
\end{eqnarray}
from which the adjoints of $\CD^{1,2}$ becomes
\begin{eqnarray}
  \langle \CD^{(1,2)^\dagger} \tilde \chi^{1,2} | \chi^{1,2} \rangle
  =
  \langle \tilde \chi^{1,2} | \CD^{(1,2)} \chi^{1,2} \rangle\ .
\end{eqnarray}
It will be shown that the redefined fermion fields $\eta^{1,2}$
\begin{eqnarray}
  \eta^1 = \frac{1}{\sqrt{1+e^{2|m_{20}|{\bf x}^3}}} \chi^1\ ,
  \qquad
  \eta^2 = \frac{1}{1+e^{2|m_{20}|{\bf x}^3}} \chi^2\ ,
\end{eqnarray}
are convenient to study their zero-modes in manifest normalizability.
Then, one can rewrite the fermion quadratic pieces in the sigma-model
Lagrangian as
\begin{eqnarray}
  \langle \chi^{1,2} | \CD^{(1,2)} \chi^{1,2} \rangle
  = \int d{\bf x}^3 \ \eta^\dagger_{1,2} D^{(1,2)} \eta^{1,2}\ .
\end{eqnarray}
One finds that the equations of motions for $\eta^{1,2}$
can be simplified as
\begin{eqnarray}
  \omega \eta^{1} &\equiv& D^{(1)} \eta^{1} = \Bigg[ i \t^3 \partial_3 -
  \hat \t_{m_{10}}  + \hat \t_{m_{20}}\bigg(\frac{|z^2|^2}{1+|z^2|^2} \bigg) \Bigg]
  \eta^{1} \nonumber \\
  \omega \eta^{2} &\equiv& D^{(2)} \eta^{2} = \Bigg[ i \t^3 \partial_3 -
  \hat \t_{m_{20}} \bigg( 1 -\frac{2|z^2|^2}{1+|z^2|^2} \bigg) \Bigg] \eta^{2}\ .
\end{eqnarray}
Inserting the explicit configuration of the kink solution, the above
differential operators can be reduced to
\begin{eqnarray}
  D^{(1)} &=& i \t^3 \partial_3  - \hat \t_{m_{10}} + \hat \t_{m_{20}} f({\bf x}^3) \big)\ , 
  \nonumber \\
  D^{(2)} &=& i \t^3 \partial_3  - \hat \t_{m_{20}} \big( 1 - 2 f({\bf x}^3) \big)
\end{eqnarray}
with
\begin{eqnarray}
  f({\bf x}^3) = \frac{e^{2|m_{20}|{\bf x}^3}}{1+ e^{2|m_{20}|{\bf x}^3}}\ ,
  \qquad
  \partial_3 f = 2|m_{20}|f (1- f) \geq 0\ .
\end{eqnarray}
\begin{figure}[t]
\begin{center}
\includegraphics[width=15cm]{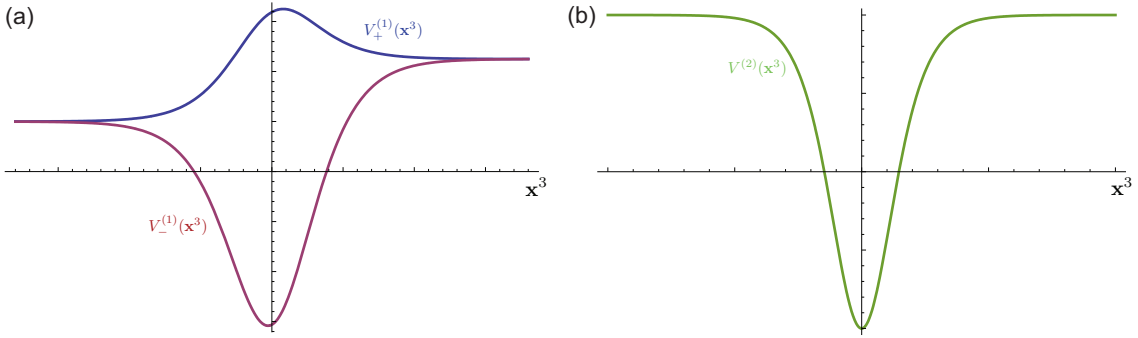}
\caption{The profiles of the effective potentials (a) $V^{(1)}_\pm({\bf x}^3)$
and (b) $V^{(2)}({\bf x}^3)$ in the case of $|m_{20}|>|m_{10}|$.}\label{potential}
\end{center}
\end{figure}

Assuming the alignment of phases of $m_{10}$ and $m_{20}$,
it is then easy to show that, for $\eta^1$,
\begin{eqnarray}
  D^{(1)\dagger} D^{(1)} = D^{(1)} D^{(1)\dagger} =
  && \hspace*{-0.5cm}
  \bigg[ -\partial_3^2 + |m_{10}|^2 - 2 |m_{10}||m_{20}| f
  + |m_{20}|^2 f^2 \bigg]{\bf 1}_4 + i \partial_3 f
  \t^3 \hat \t_{m_{20}}
%
  \nonumber \\
  \equiv && \hspace*{-0.5cm} -\partial_3^2 + V^{(1)}_\pm({\bf x}^3) \ ,
\end{eqnarray}
where the effective potentials are given by
\begin{eqnarray}
  V^{(1)}_\pm = \big(|m_{10}| - |m_{20}|f \big)^2 \pm 2 |m_{20}|^2 f (1-f)\ ,
  \qquad i\t^3\hat \t_{m_{20}} \doteq \pm |m_{20}|\ .
\end{eqnarray}
By definition, $V^{(1)}_+ \geq V^{(1)}_-$ always.
The profile of the effective potentials $V^{(1)}_\pm({\bf x}^3)$ is depicted
in figure \ref{potential} (a),
where you can see their extremum and asymptotic values are given by
\begin{eqnarray}
  V^{(1)}_{+ \text{ min}} \hspace*{-0.3cm}&=& \hspace*{-0.3cm}
  \big(|m_{20}|^2 - |m_{10}|\big)^2 + |m_{10}|^2\ ,
  \  \ \left\{
  \begin{array}{l}
  V^{(1)}_+({\bf x}^3 \to -\infty) =  |m_{10}|^2  \\
  V^{(1)}_+({\bf x}^3 \to +\infty) =  \big(|m_{20}| - |m_{10}|\big)^2
  \end{array} \right.  \\
  V^{(1)}_{- \text{ max}}\hspace*{-0.3cm}&=& \hspace*{-0.3cm}
  - \frac23 |m_{20}|\big( |m_{20}| - |m_{10}| \big)
  - \frac13 |m_{20}|^2\ ,
  \ \left\{
  \begin{array}{l}
  V^{(1)}_+({\bf x}^3 \to -\infty) = |m_{10}|^2  \\
  V^{(1)}_+({\bf x}^3 \to +\infty) = \big(|m_{20}| - |m_{10}|\big)^2
  \end{array} \right.   \nonumber
\end{eqnarray}
from which one can show that $D^{(1)}D^{(1)\dagger}$, $D^{(1)\dagger} D^{(1)}$
with $i\t^3\hat \t_{m_{20}}=+ |m_{20}|$ becomes manifestly positive definite.
It implies that there is no normalizable zero-modes for the above chirality.
For another chirality $i\t^3\hat \t_{m_{20}}=-|m_{20}|$,
one can have a normalizable zero-mode $\eta^{(1)}_0$
\begin{eqnarray}
  \eta^{1}_0 = \frac{e^{|m_{10}|{\bf x}^3}}{\sqrt{1+ e^{2 |m_{20}|{\bf x}^3}}}
  \e_0 \ , \  i\t^3\hat \t_{m_{20}} \e_0 = - |m_{20}|\e_0\  \
  \Rightarrow \ \ \chi_0^1 = e^{|m_{10}|{\bf x}^3}\ ,
\end{eqnarray}
provided that $|m_{20}| \geq |m_{10}|$.

Let us now in turn consider the Dirac operator for $\eta^2$. One can
again easily show that
\begin{eqnarray}
  D^{(2)\dagger} D^{(2)} = D^{(2)} D^{(2)\dagger} =
  && \hspace*{-0.5cm}
  \bigg[ -\partial_3^2 + |m_{20}|^2 \big( 1 - 2 f\big)^2 \bigg]{\bf 1}_4
  + \bigg[2 |m_{20}| f (1-f) \bigg]i \t^3 \hat \t_{m_{20}} \nonumber \\
  = && \hspace*{-0.5cm}
  \left\{
  \begin{array}{lcc}
    -\partial_3^2 + |m_{20}|^2 & \text{for} & i \t^3 \hat \t_{m_{20}} \doteq + |m_{20}|\\
    -\partial_3^2 + |m_{20}|^2 \big( 1- 8 f (1-f) \big)  & \text{for} &
    i \t^3 \hat \t_{m_{20}} \doteq - |m_{20}|
  \end{array}
  \right. \nonumber \\
  = && \hspace*{-0.5cm}
  \left\{
  \begin{array}{lc}
    -\partial_3^2 + |m_{20}|^2 \geq 0 &   \\
    -\partial_3^2 + V^{(2)}({\bf x}^3) &
  \end{array}
  \right.\ ,
\end{eqnarray}
which implies that there is no nomarlizable zero-modes for the
former chirality $i \t^3 \hat \t_{m_{20}} =  |m_{20}|$.
On the other hand, the effective potential $V^{(2)}({\bf x}^3)$, depicted in figure
\ref{potential} (b), has its minimum and asymptotic values like
\begin{eqnarray}
  V^{(2)}_\text{min} = - |m_{20}|^2 \ ,
  \ \
  V^{(2)} \ \to \ |m_{20}|^2 \text{ as } {\bf x}^3 \to \pm\infty\ ,
\end{eqnarray}
from which one can expect a normalizable zero-mode
$\eta^2_0$ of chirality $i \t^3 \hat \t_{m_{20}} = - |m_{20}|$
whose the explicit expression becomes
\begin{eqnarray}
  \eta^2_0 = \frac{1}{\text{cosh}\Big[ |m_{20}|{\bf x}^3 \Big]} \e_0\ \
  \Rightarrow \ \ \chi^2_0 = e^{|m_{20}|{\bf x}^3} \e_0 \ .
\end{eqnarray}
%

\subsection{the two-kink moduli space metric}
As discussed in literatures, a general kink can decompose
into several fundamental kinks. Each of fundamental kink
has two obvious collective coordinates, position and phase.
It implies that the moduli space of kinks is therefore
toric K\"ahler manifold.

For computational simplicity and concreteness,
let us consider the present model with $m_{20}=2m_{10}\equiv2m$.
From (\ref{modulimetric}), the metric components can read
\begin{eqnarray}
  g_{1\bar 1} &=&  4 \frac{|\z^2|^2}{|\z^1|^6}
  \Bigg[\frac{r}{4m} F \big( |\z^1|^4/|\z^2|^2\big) \Bigg]
  \nonumber \\
  g_{2\bar 2} &=& \frac{r}{4m} \frac{1}{|\z^2|^2} + \frac{|\z^1|^2}{|\z^1|^6}
  \Bigg[\frac{r}{4m} F \big( |\z^1|^4/|\z^2|^2\big) \Bigg]
  \nonumber \\
  g_{1\bar 2} &=& -2\frac{\bar \z_1 \z^2}{|\z^1|^6}
  \Bigg[\frac{r}{4m} F \big( |\z^1|^4/|\z^2|^2\big) \Bigg]\ ,
\end{eqnarray}
where $F(x)$ is defined in Eqs.~(\ref{F1},\ref{F2}).
%
%
%
%
Bosonic kinetic terms of interacting multi-kinks therefore
take the following form
\begin{eqnarray}
  L^\text{kin}_\text{boson} =
  L_\text{com} + L_\text{rel} \ , \nonumber
\end{eqnarray}
where
\begin{eqnarray}
  L_\text{com} = \frac{r}{4m} \Big|d \text{log} \z^2 \Big|^2\ ,
  \qquad
  L_\text{rel} = \frac{r}{4m} F \big( |\z^1|^4/|\z^2|^2\big)
  \Bigg|d \frac{\z^2}{{\z^1}^2}\Bigg|^2\ .
\end{eqnarray}
In the limit of $\frac{|\z^2|}{|\z^1|^2} \to \infty$,
$L_\text{rel}$ is asymptotic to
\begin{eqnarray}
  L_\text{rel} \simeq \frac{r}{4m} \cdot \frac{\pi}{4}
  \bigg| d \frac{\z^1}{\sqrt{\z^2}}\bigg|^2\ .
\end{eqnarray}
Note that the moduli space metric of
interacting two-kinks (, or multi kinks in
four-dimensional $\CN=2$ SQED)
has been explored by David Tong \cite{Tong:2002hi}, although our
result appears slightly different from his.

\subsection{supersymmetric low energy dynamics with potential }

For completeness, we present in this
section a short review on  supersymmetric nonlinear
sigma-model quantum mechanics with potential.
Let us begin by the Lagrangian which takes the following form
\begin{eqnarray}
  \CL_\text{kin} = \frac12 g_{IJ}
  \bigg[ \partial_0 \Phi^I \partial_0 \Phi^J
  + i \Psi^I D_0 \Psi^J \bigg] \ ,
\end{eqnarray}
where the covariant derivatives are
\begin{eqnarray}
  D_0 \Psi^I = \partial_0 \Psi^I + \partial_0 \Phi^K
  \G^I_{JK} \Psi^K \ .
\end{eqnarray}
and the fermions are real. Since the kink solitons possess
equal number of bosonic and fermionic collective coordinate,
this quantum mechanics is appropriate for the

The above Lagrangian has a real supersymmetry
whose N\"other charge is given by
\begin{eqnarray}
  \CQ = i \sqrt2  g_{IJ} \Psi^I \partial_0 \Phi^I \ .
\end{eqnarray}
Once we quantize the system. the real fermion fields $\Psi^I$
cab be represented as gamma matrices $\G^I$
\begin{eqnarray}
  \big\{ \Psi^I, \Psi^J \big\} = \d^{IJ} \ \to \
  \Psi^I \doteq \frac{1}{\sqrt2} \G^I\ .
\end{eqnarray}
It implies that the supercharge can be represented
on the Hilbert space as the spinorial Dirac operator
\begin{eqnarray}
  \CQ \doteq i \G^I \nabla_I = i \G^I \Big( \partial_I + \frac14
  {\w_I}_{AB} \G^{AB} \Big)\ .
\end{eqnarray}
When the geometry has a restricted holonomy, the
supersymmetry is enhanced. In particular, for a K\"ahler
space such as our multi-kink moduli space,
the supersymmetry is enhanced to $\CN=2$.

One may introduce to the above model
a supersymmetry-preserving deformation of the form
\begin{eqnarray}
  \CL_\text{def} = - \frac{1}{2} \Big[ g_{IJ} G^I G^J +
  i \nabla_I G_J \Psi^I \Psi^J \Big] \ .
\end{eqnarray}
One can show that the total Lagrangian $\CL=\CL_\text{kin}+
\CL_\text{def}$ is invariant under a
supersymmetry whose N\"other charge is deformed as
\begin{eqnarray}
  \CQ =\sqrt2 \Psi^I \Big[ i g_{IJ} \dot\Phi^J + G_I \Big]\ .
\end{eqnarray}
After canonical quantization,
demanding the Jacobi identity for the deformed supercharge
tells us that $G^I$ in fact turns out to be a Killing vector
field
\begin{eqnarray}
  \big[ \CQ , \big\{ \CQ, \CQ \big\} \big] = 0 \
  \to \ \nabla_{I} G_{J} + \nabla_{J} G_{I}= 0 \ .
\end{eqnarray}
When the manifold is K\"ahler with the complex structure $J$, $\CN=2$ supersymmetry
remain consistent with introduction of $G$ provided
that $G$ is not only Killing but also holomorphic,
\begin{equation}
{\cal L}_GJ=0 \,.
\end{equation}

One can split $\big\{ \CQ , \CQ \big\}$
into two conserved quantities as
\begin{eqnarray}
  \big\{ \CQ , \CQ \big\} = 4 \big( \CH - \CZ \big) \ ,
\end{eqnarray}
where $\CH$ and $\CZ$ denote Hamiltonian and central charge
\begin{eqnarray}
  \CH &=& \frac12 g_{IJ}\Big[ \partial_0 \Phi^I \partial_0 \Phi^J
  + G^I G^J \Big] + \frac i2 \nabla_I G_J \Psi^I \Psi^J \ ,
  \nonumber \\
  \CZ &=& G_I \partial_0 \Phi^I
  - \frac i2 \nabla_I G_J \Psi^I \Psi^J \ .
\end{eqnarray}
Note here that the positive energy BPS states of
real supersymmetry then preserve all the supercharges of
the moduli space dynamics.
As a final comment, the deformed supercharge
now in turn can be represented as
\begin{eqnarray}
  \CQ \doteq \G^I \big( i \nabla_I + G_I \big)
\end{eqnarray}
since we may view the wavefunctions  as sections of the
spinor bundle over the moduli space.

\end{document}